\documentclass[11pt]{article}
\usepackage{cite}
\usepackage{mdframed}
\usepackage{epsfig,subfigure}
\usepackage{amssymb}
\usepackage[fleqn]{amsmath}
\usepackage{color}
\usepackage{arydshln}
\def\QED{\mbox{\rule[0pt]{1.5ex}{1.5ex}}}

\usepackage{graphicx}
\usepackage{color}
\usepackage[dvipsnames]{xcolor}

\definecolor{armygreen}{rgb}{0.29, 0.33, 0.13}

\usepackage{amssymb}
\usepackage{amsmath,amsfonts,amssymb}
\usepackage{verbatim}
\usepackage{stfloats}
\usepackage[bookmarks=false]{}
\usepackage{algorithm}
\usepackage{algcompatible}

\newtheorem{theorem}{Theorem}
\newtheorem{corollary}{Corollary}

\newtheorem{lemma}{Lemma}

\newtheorem{remark}{Remark}

\newtheorem{example}{Example}

\newcommand\blfootnote[1]{%
  \begingroup
  \renewcommand\thefootnote{}\footnote{#1}%
  \addtocounter{footnote}{-1}%
  \endgroup
}

\begin{document}
\date{}

\title{
Secure Groupcast: 
Extra-Entropic Structure and Linear Feasibility
}
\author{\normalsize Hua Sun \\
}

\maketitle

\blfootnote{
Hua Sun (email: hua.sun@unt.edu) is with the Department of Electrical Engineering at the University of North Texas. }

\maketitle

\begin{abstract}
In the secure groupcast problem, a transmitter wants to securely groupcast a message with the maximum rate to the first $N$ of $K$ receivers by broadcasting with the minimum bandwidth, where the $K$ receivers are each equipped with a key variable from a known joint distribution. Examples are provided to prove that different instances of secure groupcast that have the same entropic structure, i.e., the same entropy for all subsets of the key variables, can have different maximum groupcast rates and different minimum broadcast bandwidth. Thus, extra-entropic structure matters for secure groupcast. 
Next, the maximum groupcast rate is explored when the key variables are generic linear combinations of a basis set of independent key symbols, i.e., the keys lie in generic subspaces. The maximum groupcast rate is characterized when the dimension of each key subspace is either small or large, i.e., the extreme regimes. For the intermediate regime, various interference alignment schemes originated from wireless interference networks, such as eigenvector based and asymptotic schemes, are shown to be useful.

\end{abstract}

\newpage

\allowdisplaybreaks
\section{Introduction}
Building an efficient secure communication network is a central problem in information theory, for which insights are obtained from studying canonical system models, e.g., ideas for secure point-to-point communication protocols emerge out of the study of Shannon's one-time pad system \cite{shannon1949}. Aiming to shed light on secure group communication protocols, we consider a recently proposed multi-user extension of the one-time pad system - secure groupcast \cite{Sun_Securegroupcast}.

In secure groupcast, a transmitter wishes to communicate a common message $W$ of $L_W$ bits to the first $N$ of $K$ receivers, i.e., the last $E = K-N$ receivers are eavesdroppers. Each receiver $k \in \{1,\cdots,K\}$ shares a key variable $Z_k$ with the transmitter. The key variables $(Z_1, \cdots, Z_K)$ are $L$ length extensions of a discrete memoryless source with a known joint distribution. The message is groupcast through broadcasting a signal $X$ of $L_X$ bits to every receiver such that combining with the known key $Z_k$, qualified Receiver $k \in \{1,\cdots, N\}$ can recover $W$ while eavesdropping Receiver $k \in \{N+1, \cdots, K\}$ learns nothing about $W$. The communication performance is measured by the secure groupcast rate $R = L_W/L$ and the broadcast bandwidth $\beta = L_X/L$. We naturally look for secure groupcast schemes with the maximum groupcast rate, termed the capacity $C$, and the minimum broadcast bandwidth $\beta^*$.

The first question studied in this work is {\em whether $C$ and $\beta^*$ can always be described by the entropy of all subsets of the key variables $(Z_1, \cdots, Z_K)$}, i.e., are entropy measures sufficient to characterize the capacity and the minimum broadcast bandwidth for secure groupcast? The answer turns out to be negative. To show this, we construct two secure groupcast instances with $N=2$ qualified receivers and $E=1$ eavesdropping receiver (i.e., $K=3$) such that the entropy of all $2^3 - 1$ subsets of $(Z_1, Z_2, Z_3)$ and the capacity $C$ are identical, while the minimum broadcast bandwidth $\beta^*$ required to achieve the capacity is different. As a result, $\beta^*$ is not a function of only the entropy measures of the key variables, and extra-entropic structure matters. Along the similar line, we construct two secure groupcast instances with $N=2$ qualified receivers and $E=2$ eavesdropping receivers (i.e., $K=4$) with the same entropy of all subsets of $(Z_1, Z_2, Z_3, Z_4)$, while the capacity $C$ is different. Therefore, $C$ cannot be expressed as a function of only the entropy measures of the key variables. The necessity of extra-entropic structures is related to one of the biggest mysteries in network information theory, i.e., the involvement of auxiliary variables in capacity characterizations. If $C, \beta^*$ may have a closed-form characterization in terms of entropy measures, we need to invoke (highly non-trivial) auxiliary variables (beyond the $K$ keys) that do not appear in the problem statement. The role of extra-entropic structures might lurk under many network information theory problems and has started to be revealed, e.g., in the context of computation broadcast \cite{Sun_Jafar_CB}.

The instances constructed in studying the first question above turn out to have {\em linear} keys, i.e., the key variables are linear combinations of a basis set of independent symbols, and the optimal achievable schemes are based on vector linear coding. This motivates us to delve deeper into the secure groupcast problem with linear keys. Another reason for linear keys and linear schemes to be interesting is that even the more preliminary form of combinatorial keys (i.e., independent uncoded keys that are shared among subsets of receivers) requires sophisticated design of the signal spaces used by the key variables $(Z_1, \cdots, Z_K)$ and the message $W$, in forming the transmit signal $X$. As such, we wish to understand the necessary and sufficient condition for a linear scheme to be feasible, i.e., the linear feasibility question. Last but not least, when the key variables are subject to design (i.e., the compound secure groupcast problem \cite{Sun_CompoundSG}), the tradeoff between key storage and broadcast bandwidth remains open and all known schemes are linear. The understanding of linear schemes will help the design of key variables. The considerations listed above lead us to the second question, where we study the natural set-up of generic linear keys.

The second question studied in this work is {\em the capacity characterization of secure groupcast with generic linear keys}, i.e., for one key block, each key variable consists of $d$ linear combinations of $m$ independent basis key symbols. Further, the linear combination coefficients are drawn independently and uniformly from a finite field. We focus on the case where the finite field is sufficiently large and study the capacity characterization for almost all cases (i.e., results that hold almost surely). In other words, the setting studied is where the key space of each receiver is a generic $d$-dimensional subspace from an $m$-dimensional overall key space. 
Interestingly, the capacity $C$ depends crucially on ratio of the key space dimension seen by each receiver to the total key dimension, i.e., $d/m$, whose reciprocal $\gamma = m/d$ can be viewed as the space expansion factor. When $\gamma$ is large, the key subspaces are far apart from each other such that the groupcast rate will be higher; when $\gamma$ is small, the key subspaces are overlapping to a large extent such that the groupcast rate will be lower. We show that for any number of qualified receivers $N$ and any number of eavesdropping receivers $E$, the capacity is $C/d = 1$ when $\gamma \geq \min(N+1, E+1)$ and $C/d = \gamma - 1$ when $\gamma \leq \max(1+1/N, 1+1/E)$. When either $N = 1$ (secure unicast) or $E = 1$ (secure multicast), the capacity is characterized for all possible $\gamma$. This result is generalized along two lines. For the first line, we consider the simplest uncovered setting where $N=2, E=2$, and provide partial characterization for the remaining regimes of $3/2 < \gamma < 3$. For the second line, we fix $\gamma = 2$ (i.e., each key subspace has half the dimension of the overall key space), and add more qualified or eavesdropping receivers to the basic $N=2, E=2$ system. In particular, we show that if we add one qualified or eavesdropping receiver, i.e., set $N = 3$ or $E = 3$, then the earliest known interference alignment schemes, first appeared in the $2$ user wireless $X$ network \cite{Jafar_Shamai}, are useful; if we add $2$ qualified or eavesdropping receivers, i.e., set $N=4$ or $E = 4$, then eigenvector based interference alignment schemes, originated from the $3$ user wireless interference network \cite{Cadambe_Jafar_int}, can be applied; if we further add more qualified receivers or eavesdropping receivers, i.e., set $N > 4$ or $E > 4$, then asymptotic interference alignment schemes, which lie in the core of the canonical half-the-cake result for wireless interference networks \cite{Cadambe_Jafar_int, Jafar_FnT}, play significant roles. Ideas that resemble the wireless counterpart appear in secure groupcast, e.g., spatial normalization, diagonal channel coefficients, over-constrained linear systems, duality, and space overlaps. We will go through these ideas in the following sections of this work.

\bigskip
{\it Notation: For positive integers $K_1, K_2, K_1 \leq K_2$, we use the notation $[K_1:K_2] = \{K_1, K_1+1,\cdots, K_2\}$. Define the notation $Z_{K_1:K_2}$ as the vector $(Z_{K_1}; \cdots; Z_{K_2})$ if $K_1 \leq K_2$ and as the null vector otherwise.
The notation $|\mathcal{Q}|$ is used to denote the cardinality of a set $\mathcal{Q}$. 
We use ${\bf 0}$ to denote a matrix whose each element is $0$ and use ${\bf I}_{d\times d}$ to denote the identity matrix of dimension $d$. For a matrix ${\bf H}$, ${\bf H}(i:j, :)$ is used to denote the sub-matrix of ${\bf H}$ formed by retaining only the $i$-th row to the $j$-th row.
}


\section{Problem Statement}\label{sec:model}
Consider $K$ discrete random variables $z_1, \cdots, z_K$ of finite cardinality, drawn from an arbitrary joint distribution $P_{z_1,\cdots, z_K}$. In this work, we focus on the linear setting, where $z_k, k \in [1:K]$ are arbitrary linear combinations of a basis set of independent symbols from a finite field. Let the basis symbols be specified through the $m \times 1$ column vector ${\bf s} = (s_1; \cdots; s_m) = s_{1:m}$, where $s_i, i \in [1:m]$ are i.i.d. uniform symbols from a finite field $\mathbb{F}_p$ for a prime power $p$. Since all variables $z_k$ are linear combinations of the basis symbols, they are represented by $1\times m$ vectors of linear combining coefficients.
Each variable $z_k$ is then specified in terms of such vectors, $z_k = {\bf H}_k {\bf s}, $ where ${\bf H}_k \in \mathbb{F}_p^{d\times m}$, and each $z_k$ contains $d$ symbols from $\mathbb{F}_p$, where each symbol is a linear combination of the basis symbols with coefficients specified by one row vector of ${\bf H}_k$. $Z_1, \cdots, Z_K$ are $L$ length extensions of $z_1, \cdots, z_K$, where each block $Z_1(l), \cdots, Z_K(l)$ is produced i.i.d. according to $P_{z_1, \cdots, z_K}$.

The secure groupcast problem is comprised of a transmitter and $K$ receivers. The key variable $Z_k$ is shared between the transmitter and Receiver $k$. The transmitter wishes to send a message $W$ that has $L_W$ i.i.d. uniform symbols from $\mathbb{F}_p$ and is independent of the key variables $Z_1, \cdots, Z_K$ to the first $N < K$ receivers.
\begin{eqnarray}
&& H(W) = L_W ~\mbox{(in $p$-ary units)}, \label{h1} \\
&& I(W; Z_1, \cdots, Z_K) = 0. \label{h2}
\end{eqnarray}
To securely groupcast the message $W$, the transmitter broadcasts a signal $X$ of $L_X$ symbols from $\mathbb{F}_p$ to every receiver. Each qualified receiver can decode $W$ with no error\footnote{For the linear key setting, all achievable schemes of this work (except that in Section \ref{sec:eleak}) have zero error and zero leakage. 
Note that all converse results of this work also hold under $\epsilon$ error and $\epsilon$ leakage.}.
\begin{eqnarray}
\mbox{[Correctness]} ~~ H(W | X, Z_k) = 0, \forall k \in [1:N]. \label{corr}
\end{eqnarray}
Each unqualified (eavesdropping) receiver learns no information about $W$.
\begin{eqnarray}
\mbox{[Security]} ~~ I(W; X, Z_k) = 0, \forall k \in [N+1: K]. \label{sec}
\end{eqnarray}
The secure groupcast rate characterizes how many symbols of the message are securely groupcast per key block and the broadcast bandwidth characterizes how many symbols of the transmit signal are broadcast per key block to securely groupcast a message of certain rate.
\begin{eqnarray}
R = \frac{L_W}{L}, ~\beta(R) = \frac{L_X}{L}. \label{rate}
\end{eqnarray}
A rate $R$ is said to be achievable if there exists a secure groupcast scheme (that satisfies the correctness constraint (\ref{corr}) and the security constraint (\ref{sec})) of rate greater than or equal to $R$. The supremum of achievable rates is called the capacity $C$. A broadcast bandwidth $\beta(R)$ is said to be achievable if there exists a secure groupcast scheme of rate greater than or equal to $R$ and of broadcast bandwidth smaller than or equal to $\beta(R)$. The infimum of achievable broadcast bandwidth is called the minimum broadcast bandwidth $\beta^*(R)$.

\subsection{Preliminary Result}
We recall a useful converse result on $R$ and $\beta(R)$ that is stated in the following theorem and will be used later. The proof can be found in \cite{Sun_Securegroupcast}.
\begin{theorem}\label{thm:con}
{\normalfont(Theorem 1 and Theorem 2 in \cite{Sun_Securegroupcast})}
For the secure groupcast problem, 
we have
\begin{eqnarray}
&& R \leq H(z_q | z_e), \forall q \in [1:N], \forall e \in [N+1:K], \label{con:rate} \\
&& \beta(R) \geq I(X; W, Z_{1:K} | U_e)/L \geq |\mathcal{Q}| R - \Big(\sum_{i=1}^{|\mathcal{Q}|}H(z_{q_i} | u_e) - H(z_{q_1}, \cdots, z_{q_{|\mathcal{Q}|}}| u_e) \Big),\notag\\
&&~\forall \mathcal{Q} = \{q_1, \cdots, q_{|\mathcal{Q}|}\} \subset [1:N], \forall e \in [N+1:K], \forall u_e ~\mbox{s.t.}~ H(u_e | z_e) = 0. \label{con:beta}
\end{eqnarray}
\end{theorem}


\section{Extra-Entropic Structure}
In this section, we consider the question if the capacity $C$ and the minimum broadcast bandwidth for capacity achieving schemes $\beta^*(C)$ can be characterized by the entropy of all subsets of the key variables. Our result shows that the answer is no in general.

\begin{theorem}\label{thm:extra}
There exist instances of the secure groupcast problem where the entropy structure of the key variables and the capacity $C$ are the same, while the minimum broadcast bandwidth $\beta^*(C)$ is different; there exist instances of the secure groupcast problem where the entropy structure of the key variables is the same while the capacity values $C$ are different. Thus extra-entropic structure matters for secure groupcast.
\end{theorem}

{\it Proof:} First, consider $\beta^*(C)$. We present two instances of secure groupcast, say $\mbox{SG}_1, \mbox{SG}_2$, that have the same entropy values of all subsets of the key variables and the same capacity. Yet, these two instances have different minimum broadcast bandwidth for capacity achieving schemes. Incidentally, both instances have linear keys and are specified as follows. For both instances\footnote{The parameters $N=2, K=3$ are the smallest so that the instances are the simplest, because if $N=1$ (single qualified receiver, i.e., the secure unicast setting), then $\beta^*(C) = \min_{e\in[2:K]} H(z_1|z_e)$ is fully characterized by the entropy structure of the key variables (see Theorem 9 in \cite{Sun_Securegroupcast}).}, $N=2, K=3$ and each key variable consists of $d = 2$ linear combinations of $m = 3$ basis key symbols $s_1, s_2, s_3$ from any field $\mathbb{F}_p$.
\begin{eqnarray}
\begin{array}{llllll}
\mbox{SG}_1: & z_1 =  (s_1; s_2), & z_2 = (s_1; s_3), & z_3 = (s_2; s_3);\\
\mbox{SG}_2: & z_1 = (s_1; s_2), & z_2 = (s_1; s_3), & z_3 = (s_1; s_2 + s_3).
\end{array}
\end{eqnarray}
The entropy values of all subsets of $(z_1, z_2, z_3)$ are found as follows.
\begin{eqnarray}
H(z_i) = 2, \forall i \in \{1,2,3\}, H(z_i, z_j) = H(z_1, z_2, z_3) = 3, \forall i, j \in \{1,2,3\}, i \neq j.
\end{eqnarray}
So the entropy structure of both $\mbox{SG}_1$ and $\mbox{SG}_2$ is the same. We next characterize the capacity and the minimum broadcast bandwidth for both instances. 
\begin{eqnarray}
\mbox{SG}_1: && C_{\mbox{\tiny\it SG}_1} = 1, ~\beta_{\mbox{\tiny\it SG}_1}^*(C) = 1. \\
&& \mbox{Rate Converse:}~~~~~~~~~ R \leq H(z_1 | z_3) = 1 ~~(\mbox{set $q=1, e= 3$ in (\ref{con:rate}) of Theorem \ref{thm:con}}). \\
&& \mbox{Bandwidth Converse:}~\beta(C) \geq C = 1 ~~~~~~~(\mbox{set $\mathcal{Q} = \{1\}, u_e = ()$ in (\ref{con:beta})}). \\
&& \mbox{Achievability:}~~~~~~~~~~~ X = W + s_1. \\
\mbox{SG}_2: && C_{\mbox{\tiny\it SG}_2} = 1, ~\beta_{\mbox{\tiny\it SG}_2}^*(C) = 2. \\
&& \mbox{Rate Converse:}~~~~~~~~~ R \leq H(z_1 | z_3) = 1 ~~(\mbox{set $q=1, e= 3$ in (\ref{con:rate}) of Theorem \ref{thm:con}}). \\
&& \mbox{Bandwidth Converse:}~\beta(C) \geq 2C - (H(z_1|s_1) + H(z_2|s_1) - H(z_1, z_2 | s_1)) \\
&&~~~~~~~~~~~~~~~~~~~~~~~~~~~~~~~~~~~~ = 2 ~~~~~~~~~~~~~(\mbox{set $\mathcal{Q} = \{1,2\}, e = 3, u_e = s_1$ in (\ref{con:beta})}). \\
&& \mbox{Achievability:}~~~~~~~~~~~ X = (W + s_2; -W + s_3).
\end{eqnarray}
Note that we use $L=1$ key block so that $Z_i = z_i$ and $W$ has $L_W = 1$ symbol. Therefore while the capacity for $\mbox{SG}_1, \mbox{SG}_2$ is the same, the minimum broadcast bandwidth is different. A closer look at the proof reveals that the converse bound (\ref{con:beta}) has an auxiliary variable $u_e$ that might need to be set differently for different secure groupcast instances.

Second, consider $C$. We present two instances of secure groupcast, say $\mbox{SG}_3, \mbox{SG}_4$, that have the same entropy values of all subsets of the key variables. Yet, these two instances have different capacity values. For both instances\footnote{$N=2, K=4$ is the simplest setting, because if either $N=1$ or $K-N=1$ (secure unicast or secure multicast), then the capacity is fully characterized by the entropy structure of the key variables (see Theorem 9 in \cite{Sun_Securegroupcast}).}, $N=2, K=4$, the keys are linear, and each key variable consists of $d = 9$ linear combinations of $m = 15$ basis key symbols $s_1, \cdots, s_{15}$ from field $\mathbb{F}_p$, where $p \geq 5$ is a prime.
\begin{eqnarray}
\begin{array}{llllll}
\mbox{SG}_3: & z_1 =  (s_{1:3}; s_{4:6}; s_{7:9}) \\
& z_2 = (s_{1:3}; s_{10:12}; s_{13:15}) \\
& z_3 = (s_{4:6}; s_{10:12}; s_{1:3}+s_{4:6}+s_{7:9}+s_{10:12}+s_{13:15}) \\ 
& z_4 = (s_{7:9}; s_{13:15}; s_{1:3}+s_{4:6}+s_{7:9}+s_{10:12}+s_{13:15}) \\
\mbox{SG}_4: & z_1 = (s_{1:3}; s_{4:6}; s_{10:12}) \\
& z_2 = (s_{1:3}; s_{7:9}; s_{13:15}) \\
& z_3 = (s_{1:3}; s_{4} + s_{7}; s_{5} + s_{8}; s_{6} + s_{9} ; s_{10} + s_{13}; s_{11} + s_{14}; s_{12} + s_{15}) \\
& z_4 =  (s_{1:3}; s_{4} + 2s_{7}; s_{5} + 3s_{8}; s_{6} + 4s_{9} ; s_{10} + 2s_{13}; s_{11} + 3s_{14}; s_{12} + 4s_{15}) .
\end{array}
\end{eqnarray}
The entropy values of all subsets of $(z_1, z_2, z_3, z_4)$ are found as follows.
\begin{eqnarray}
&& H(z_i) = 9, \forall i \in \{1,2,3,4\}, \\
&& H(z_i, z_j) = H(z_i,z_j,z_k) = H(z_1, z_2, z_3, z_4) = 15, \forall i, j, k \in \{1,2,3,4\}, i \neq j. 
\end{eqnarray}
So the entropy structure of both $\mbox{SG}_3$ and $\mbox{SG}_4$ is the same. The capacity of $\mbox{SG}_3$ is characterized as follows. $W = (W_{1:3}; W_{4:6})$ has $L_W = 6$ symbols and we use $L=1$ key block.
\begin{eqnarray}
\begin{array}{llllll}
\mbox{SG}_3: & C_{\mbox{\tiny\it SG}_3} = 6. \\
& \mbox{Converse:} & R \leq H(z_1 | z_3) = 6 ~~(\mbox{set $q=1, e= 3$ in (\ref{con:rate}) of Theorem \ref{thm:con}}). \\
& \mbox{Achievability:} & 
X = \left( \begin{array}{c}
W_{1:3} + s_{1:3} \\
W_{4:6} + s_{4:6} + s_{7:9} \\
- W_{1:3} - W_{4:6} + s_{10:12} + s_{13:15}
\end{array} \right) . \\
\end{array}
\end{eqnarray}
Correctness and security can be easily verified.
The capacity of $\mbox{SG}_4$ is more involved (e.g., the converse from Theorem \ref{thm:con} no longer suffices) and the result is presented in the following lemma.
\begin{lemma}\label{lemma:sp4}
For the secure groupcast instance $\mbox{SG}_4$, the capacity is $C_{\mbox{\tiny\it SG}_4} = 4$.
\end{lemma}

The proof is deferred to Section \ref{sec:sp4} and an outline is given here. The symbols $s_{1:3}$ are useless as all receivers know them. The remaining $6$ key symbols for each receiver can be divided into $2$ groups, and each group is essentially a generic secure groupcast instance, where each receiver has a $3$-dimensional key subspace in general position of a $6$-dimensional space, e.g., for one group, Receiver $1$ has $s_{4:6}$, Receiver $2$ has $s_{7:9}$, Receiver $3$ has $(s_{4} + s_{7}; s_{5} + s_{8}; s_{6} + s_{9})$, and Receiver $4$ has $(s_{4} + 2s_{7}; s_{5} + 3s_{8}; s_{6} + 4s_{9})$. This generic secure groupcast instance will be settled in Theorem \ref{thm:22} and the capacity is $2$ so that for $2$ groups with independent keys, the capacity of $\mbox{SG}_4$ is $4$. The insights of Theorem \ref{thm:22} can be generalized to produce the proof of Lemma \ref{lemma:sp4} (see Section \ref{sec:sp4}).

Therefore while the entropy structure of $\mbox{SG}_3, \mbox{SG}_4$ is the same, the capacity values are different. Extra-entropic structure matters and the proof of Theorem \ref{thm:extra} is complete.

\hfill\QED

\section{Linear Feasibility}
In this section, we characterize the feasibility condition of a linear secure groupcast scheme under the linear key setting. Note that while the achievable rates defined in the problem statement section are not restricted to linear schemes, linear schemes are of interest because on the one hand they are simple, and on the other hand they often turn out to be optimal for linear keys.

\bigskip
\noindent {\bf Linear Scheme:} {\it For a linear secure groupcast scheme with linear keys, $z_k = {\bf H}_k {\bf s}, {\bf H}_k \in \mathbb{F}_p^{d \times m}, {\bf s} \in \mathbb{F}_p^{m \times 1}$, the transmit signal 
\begin{eqnarray}
X = {\bf V}_W W + {\bf V} {\bf s}, ~{\bf V}_W \in \mathbb{F}_p^{L_X \times L_W}, W \in \mathbb{F}_p^{L_W \times 1}, {\bf V} \in \mathbb{F}_p^{L_X \times m} \label{eq:linear}
\end{eqnarray}
is specified by two full rank\footnote{We assume without loss of generality that ${\bf V} \in \mathbb{F}_p^{L_X \times m}$ has full row rank, i.e., $\mbox{rank}({\bf V}) = L_X \leq m$. Otherwise, some row of ${\bf V}$ is a linear combination of other rows and due to the security constraint (\ref{sec}), the corresponding linear combinations of ${\bf V}_W$ must be zero. As a result, some row of the transmit signal $X$ is a linear combination of other rows, i.e., $X$ contains some redundant row that does not need to be sent.} precoding matrices, ${\bf V}_W$ for the message $W$ and ${\bf V}$ for the key variables ${\bf s}$ such that the following properties are satisfied.
\begin{itemize}
\item Identify the overlap of the key space of $X$ and the key space of  Receiver $k \in [1:K]$, i.e., find matrices ${\bf P}_k$ and ${\bf U}_k$ of the maximum rank (if exist) such that\footnote{Equivalently, $\mbox{rank}({\bf U}_k) = \mbox{rank}({\bf P}_k) = \mbox{dim}(\mbox{rowspan}({\bf V}) \cap \mbox{rowspan}({\bf H}_k))$.}
\begin{eqnarray}
{\bf U}_k {\bf V} = {\bf P}_k {\bf H}_k, \label{eq:proj}
\end{eqnarray}
then the projection of $W$ in $X$ to the ${\bf U}_k$ space satisfies
\begin{eqnarray}
\mbox{[Correctness]} && \mbox{rank}({\bf U}_k {\bf V}_W) = L_W, ~\forall k \in [1:N], \label{eq:corr}\\
\mbox{[Security]} && {\bf U}_k {\bf V}_W = {\bf 0}, ~\forall k \in [N+1:K]. \label{eq:sec}
\end{eqnarray}
\end{itemize}
The rate achieved is $R = L_W$ and the broadcast bandwidth achieved is $\beta(R) = L_X$ as $L = 1$. Generalizations to $L > 1$ are immediate.
}
\bigskip

Note that the precoding matrices ${\bf V}_W$, ${\bf V}$ and the key matrices ${\bf H}_k$ are constants and are assumed globally known to the transmitter and all receivers, so that it is straightforward to find the projection matrices ${\bf U}_k, {\bf P}_k$ and verify the feasibility condition.
We show that the correctness constraint (\ref{eq:corr}) and the security constraint (\ref{eq:sec}) for linear schemes implies the entropic versions (\ref{corr}) and (\ref{sec}). For correctness, we have
\begin{eqnarray}
&& {\bf U}_k X = {\bf U}_k {\bf V}_W W + {\bf U}_k {\bf V} {\bf s} \overset{(\ref{eq:proj})}{=} {\bf U}_k {\bf V}_W W + {\bf P}_k {\bf H}_k {\bf s} \\
&\Rightarrow& {\bf U}_k X - {\bf P}_k {\bf H}_k {\bf s} = {\bf U}_k {\bf V}_W W \overset{(\ref{eq:corr})}{\longleftrightarrow} W, ~\forall k \in [1:N]\\
&\Rightarrow& H(W | X, Z_k) = 0
\end{eqnarray}
where ${\bf A} \longleftrightarrow {\bf B}$ means that the two matrices ${\bf A}, {\bf B}$ are invertible. For security, we use the fact that ${\bf P}_k {\bf H}_k$ contains all row vectors of ${\bf H}_k$ that can be expressed as linear combinations of the row vectors of ${\bf V}$ (i.e., all overlaps) so that the row space of the remaining vectors is orthogonal to the row space of ${\bf V}$.
\begin{eqnarray}
{\bf H}_k \longleftrightarrow ({\bf P}_k {\bf H}_k; {\bf Q}_k {\bf H}_k) \Rightarrow ~\mbox{rowspan(${\bf Q}_k {\bf H}_k$) is independent of rowspan(${\bf V}$)}. \label{eq:orth}
\end{eqnarray}
Then we have $\forall k \in [N+1:K]$
\begin{eqnarray}
I(W; X, Z_k) &\overset{(\ref{h2})}{=}& I(W; X, {\bf P}_k {\bf H}_k {\bf s} \mid {\bf Q}_k {\bf H}_k {\bf s})\\
&\overset{(\ref{eq:sec})}{=}& I(W; X \mid {\bf Q}_k {\bf H}_k {\bf s}) \\
&=& H(X \mid {\bf Q}_k {\bf H}_k {\bf s} ) - H(X \mid W, {\bf Q}_k {\bf H}_k {\bf s}) \\
&\overset{(\ref{h2})}{\leq}& L_X - H({\bf V} {\bf s} \mid {\bf Q}_k {\bf H}_k {\bf s}) \\
&\overset{(\ref{eq:orth})}{=}& L_X - H({\bf V} {\bf s}) = 0 \label{eq:e0}
\end{eqnarray}
where in the last step, we use the fact that ${\bf V} \in \mathbb{F}_p^{L_X\times m}$ has full row rank, i.e., $H({\bf V} {\bf s}) = \mbox{rank}({\bf V}) = L_X$. From now on, we will employ the simplified correctness and security constraints (\ref{eq:corr}), (\ref{eq:sec}) for achievability proofs of linear schemes.

Evidently, the conditions (\ref{eq:corr}) and (\ref{eq:sec}) are necessary as otherwise, either the qualified receiver cannot decode the desired message (refer to (\ref{eq:corr})) or the eavesdropping receiver can obtain some linear combination of the message symbols (refer to (\ref{eq:sec})).


\section{Generic Secure Groupcast}
In this section, we study the secure groupcast problem when the keys are generic linear combinations of the basis symbols, abbreviated as generic secure groupcast. In particular, $z_k = {\bf H}_k {\bf s}$, where
\begin{eqnarray}
\mbox{each element of ${\bf H}_k \in \mathbb{F}_p^{d \times m}$ is drawn independently and uniformly from $\mathbb{F}_p$ for a large $p$.} \label{eq:generic}
\end{eqnarray}
Define $\gamma = m/d \in \mathbb{Q}$ as the ratio of the dimension of the overall key space to the dimension of the generic key subspace seen by each receiver. It is convenient to adopt the normalization of the groupcast rate by $d$, $R/d$ as the rate measure. We denote the number of eavesdropping receivers by $E = K-N$, to simplify the notations. The maximum normalized rate is characterized when $\gamma$ is either small or large, in the following theorem.

\begin{theorem}\label{thm:generic}
For generic secure groupcast with $N$ qualified receivers and $E$ eavesdropping receivers, when the key of each receiver consists of $d$ generic linear combinations of $m = \gamma d \geq d$ basis symbols, the capacity is
\begin{eqnarray}
C/d = 1, && ~\mbox{when}~ \gamma \geq \min(N+1, E+1);\\
C/d = \gamma - 1, && ~\mbox{when}~ 1 \leq \gamma \leq \max(1+1/N, 1+1/E)
\end{eqnarray}
almost surely.
\end{theorem}

\begin{figure}[h]
\begin{center}
\includegraphics[width=3.3 in]{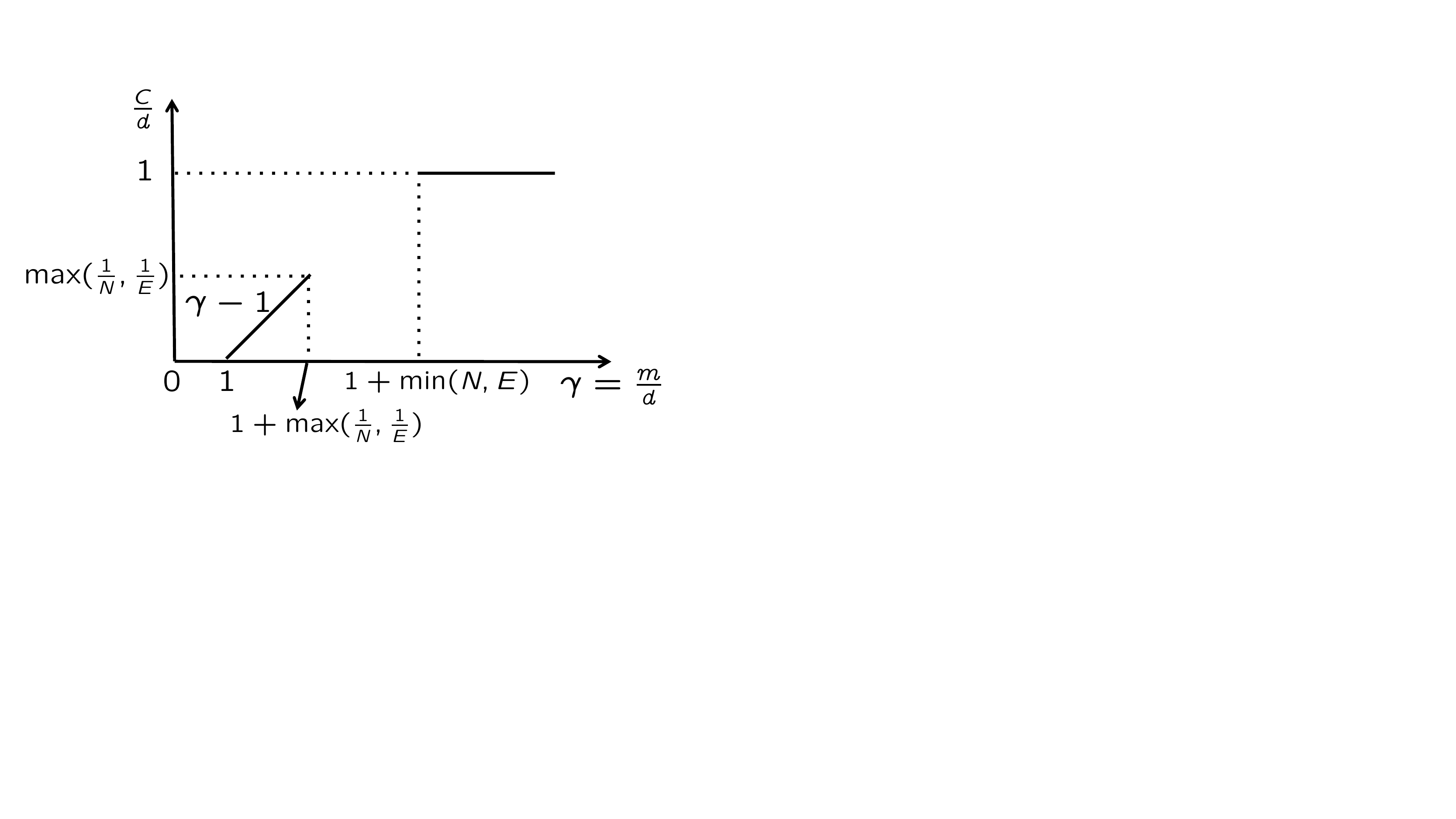}
\caption{\small Normalized generic secure groupcast capacity $C/d$ as a function of space expansion factor $\gamma = m/d$.}
\label{fig:generic}
\end{center}
\end{figure}

The result in Theorem \ref{thm:generic} is plotted in Fig.~\ref{fig:generic}. The detailed proof of Theorem \ref{thm:generic} is presented in Section \ref{sec:generic}. To illustrate the idea in a simpler setting, we give a few examples here.

\begin{example}\label{ex:g1}
(Large $\gamma$)
Suppose we have $N=2$ qualified receivers and $E=3$ eavesdropping receivers. From Theorem \ref{thm:generic}, we know that as long as $\gamma \geq \min(3,4) = 3$, then $C/d = 1$. Suppose $\gamma = 3$, e.g., $d = 1, m = \gamma d = 3$ so that each receiver has $d = 1$ generic linear combination of $m=3$ basis key symbols $s_1, s_2, s_3$, as the key $z_k$. We show that the capacity is $C = d = 1$. Converse follows immediately from (\ref{con:rate}) in Theorem \ref{thm:con}, $R \leq H(z_q | z_e) = 1, \forall q \in \{1,2\}, \forall e \in \{3,4,5\}$. Note that $z_q, z_e$ each lies in a $1$-dimensional subspace in general position of a $3$-dimensional space so that $z_q, z_e$ are linearly independent almost surely. Achievability is proved as follows. We simply send the sum of the message symbol and each key of qualified receivers, i.e., the transmit signal is set as
\begin{eqnarray}
X = ( W + z_1; W + z_2)
\end{eqnarray}
where $X \in \mathbb{F}_p^{2 \times 1}, W \in \mathbb{F}_p$. Correctness is easy to see - referring to (\ref{eq:corr}), the overlap of the key space of $X$ and the key space of qualified Receiver $q$ is $z_q$ and the projection of the message space to $z_q$ is $W$, from which $W$ can be decoded with no error. Security is guaranteed, because referring to (\ref{eq:sec}), the overlap of the key space of $X$ and the key space of eavesdropping Receiver $e$ is null because the key in $X$ has $2$ dimensions in general position, which is independent of the $1$-dimensional key space $z_e$ almost surely. In other words, the messages are sent along the qualified key spaces, which are independent of each eavesdropping key space almost surely.

The above case satisfies $N \leq E$ and when $N > E$, a different idea is required. Suppose $N = 3, E = 2, \gamma = \min(4,3) = 3$. Similarly, suppose $d = 1, m = 3$, i.e., each key space is a $1$-dimensional generic subspace of a $3$-dimensional space. We show that $C = d = 1$. The converse proof is same as above and we consider achievability. The transmit signal is designed as
\begin{eqnarray}
X = {\bf V}_W W + {\bf s}
\end{eqnarray}
where $X, {\bf V}_W, {\bf s} \in \mathbb{F}_p^{3\times 1}, W \in \mathbb{F}_p$ and ${\bf V}_W$ is chosen so that
\begin{eqnarray}
\left[ \begin{array} {c}
{\bf H}_4 \\
{\bf H}_5
\end{array} \right]_{2 \times 3}
{\bf V}_W = {\bf 0}_{2 \times 1}, \label{eq:g2}
\end{eqnarray}
i.e., ${\bf V}_W$ lies in the right null space of each eavesdropping key space. Such a ($1$-dimensional) null space exists because the overall key space has $3$ dimensions and the $2$ eavesdropping receivers see a generic $2$-dimensional subspace collectively. Correctness constraint (\ref{eq:corr}) holds, because $X$ uses the full key space such that its overlap with each qualified key space is $z_q$, and the projection of the message space ${\bf V}_W W$ to $z_q$ is not zero almost surely, i.e., ${\bf H}_{q} {\bf V}_W \neq 0, \forall q \in \{1,2,3\}$ (note that ${\bf V}_W$ is determined fully by the eavesdropping key space and is independent of the qualified key space almost surely). Security constraint (\ref{eq:sec}) holds because of the design of ${\bf V}_W$ (refer to (\ref{eq:g2})). To sum up, the message is sent along the null space of the eavesdropping key spaces, whose projection to each qualified key space is not null almost surely.

Finally, we note that the idea of the achievable scheme for the above two cases are similar to that for the minimum key storage extreme point of the compound secure groupcast problem \cite{Sun_CompoundSG}.
\end{example}

\begin{example}\label{ex:g2}
(Small $\gamma$)
Similar to the large $\gamma$ regime, we also have $2$ cases for the small $\gamma$ regime, depending on $N \leq E$ or $N > E$. The $2$ cases require different ideas and are considered sequentially.

First, suppose $N = 2, E = 3$. Theorem \ref{thm:generic} states that if $\gamma \leq \max(1+1/2, 1+1/3) = 3/2$, then $C/d = \gamma - 1$. To illustrate this, suppose $\gamma = 3/2$, e.g., $d = 2, m = \gamma d = 3$ so that each $z_k$ consists of $2$ generic linear combinations of $3$ basis key symbols $s_1, s_2, s_3$. We show that the capacity is $C = d(\gamma-1) = m-d = 1$. Converse follows from (\ref{con:rate}) in Theorem \ref{thm:con}, $R \leq H(z_q | z_e) = H(z_q, z_e) - H(z_e) = H(s_1, s_2, s_3) - H(z_e) = 3 - 2 = 1, \forall q \in \{1,2\}, \forall e \in \{3,4,5\}$. Note that $z_q, z_e$ each lies in a $2$-dimensional subspace in general position of a $3$-dimensional space so that they have full rank collectively almost surely. 
Achievability follows from the fact that the $2$ generic $2$-dimensional qualified key spaces $z_1, z_2$ have $1$-dimensional overlap in the $3$-dimensional overall key space with high probability. Denote this row vector as ${\bf H}_{\mathcal{Q}}$ so that 
\begin{eqnarray}
\mbox{rowspan}({\bf H}_{\mathcal{Q}}) = \mbox{rowspan}({\bf H}_1)\cap\mbox{rowspan}({\bf H}_2). \label{eq:g1}
\end{eqnarray}
Then the transmit signal is set as
\begin{eqnarray}
X = W + {\bf H}_{\mathcal{Q}} {\bf s}
\end{eqnarray}
where $X, W \in \mathbb{F}_p, {\bf H}_{\mathcal{Q}} \in \mathbb{F}_p^{1\times 3}, {\bf s} \in \mathbb{F}_p^{3\times 1}$. Correctness constraint (\ref{eq:corr}) follows from the construction that ${\bf H}_{\mathcal{Q}}$ lies in the key space of each qualified receiver (see (\ref{eq:g1})), so ${\bf H}_{\mathcal{Q}} {\bf s}$ and $W$ are recoverable. Security constraint (\ref{eq:sec}) follows from the observation that the row vector ${\bf H}_{\mathcal{Q}}$ is determined fully by the qualified key spaces such that it is almost surely independent of each $2$-dimensional eavesdropping generic key space in the $3$-dimensional overall key space. As a recap, the qualified key spaces have a common overlap that is independent of each eavesdropping key space with high probability and this overlap is used as the common key to send the desired message with one-time pad.

Second, suppose $N=3, E = 2$. We set $\gamma = \max(1+1/3, 1+1/2) = 3/2$, same as above. Similarly, suppose $d = 2, m = 3$, i.e., each key space is a $2$-dimensional generic subspace of a $3$-dimensional space. We show that $C = d(\gamma - 1) = 1$. The converse proof is same as above while achievability requires a somewhat dual idea. The $2$ eavesdropping key spaces $z_4, z_5$ each has $2$ dimensions and have a $1$-dimensional overlap in the $3$-dimensional overall key space almost surely. Denote this row vector as ${\bf H}_{\mathcal{E}}$ so that
\begin{eqnarray}
\mbox{rowspan}({\bf H}_{\mathcal{E}}) = \mbox{rowspan}({\bf H}_4)\cap\mbox{rowspan}({\bf H}_5).
\end{eqnarray}
Then the transmit signal is set as
\begin{eqnarray}
X = ( {\bf H}_{\mathcal{E}} \hspace{0.02in} {\bf s} ; ~W + {\bf H}_{\mbox{\scriptsize rand}} \hspace{0.03in} {\bf s})
\end{eqnarray}
where $X \in \mathbb{F}_p^{2 \times 1}, W \in \mathbb{F}_p, {\bf H}_{\mathcal{E}}, {\bf H}_{\mbox{\scriptsize rand}} \in \mathbb{F}_p^{1\times 3}, {\bf s} \in \mathbb{F}_p^{3 \times 1}$ and ${\bf H}_{\mbox{\scriptsize rand}}$ is a random row vector where each element is drawn independently and uniformly from $\mathbb{F}_p$. To verify correctness constraint (\ref{eq:corr}), note that the row vector ${\bf H}_{\mathcal{E}}$ is linearly independent of each $2$-dimensional qualified key space in the $3$-dimensional overall key space almost surely, so from $({\bf H}_{\mathcal{E}} \hspace{0.02in} {\bf s}; {\bf H}_q {\bf s}), \forall q \in \{1,2,3\}$, each qualified receiver has $3$ generic linear combinations of all basis symbols ${\bf s}$ and can fully recover ${\bf s}$. Then ${\bf H}_{\mbox{\scriptsize rand}} \hspace{0.03in} {\bf s}$ can be obtained and then $W$ is decoded with no error. To verify security constraint (\ref{eq:sec}), note that ${\bf H}_{\mathcal{E}} \hspace{0.02in} {\bf s}$ is known to each eavesdropping receiver such that no additional information is revealed and the row vector ${\bf H}_{\mbox{\scriptsize rand}}$ is linearly independent of each $2$-dimensional eavesdropping key space in the $3$-dimensional overall key space almost surely. To sum up, the message is sent along random row vectors and the common overlap of eavesdropping key spaces is broadcast to enable qualified receiver to recover the key along the random precoding vectors (that are mixed with the message) and ensure eavesdropping receiver learns no information about the message.

Finally, we note that the idea of the achievable scheme for the above two cases are similar to that for the minimum broadcast bandwidth extreme point of the compound secure groupcast problem \cite{Sun_CompoundSG}.
\end{example}

Note that when $N=1$ or $E=1$, there is no gap between the $\gamma$ regimes in Theorem \ref{thm:generic} so that the capacity is fully characterized for all $\gamma$ values. This result is stated in the following corollary.
\begin{corollary}
For generic secure unicast $(N=1)$ and generic secure multicast ($E=1$), 
the capacity is
$C/d = 1, \mbox{if}~ \gamma \geq 2$, and $C/d = \gamma - 1, \mbox{otherwise}~ 1\leq \gamma \leq 2$ almost surely.
\end{corollary}

\subsection{$N=E=2$ and Spatial Normalization}
As the settings where either $N=1$ or $E=1$ are fully understood, we proceed to consider the simplest open generic secure groupcast problem with $N=2$ and $E=2$. We start by introducing the metric - spatial normalized rate and capacity. 

The achievable spatial normalized rate of generic secure groupcast, denoted by $\overline{R}(\gamma)$, is defined as $R/d$ if the secure groupcast rate $R$ is achievable when each key $z_k  = {\bf H}_k {\bf s}$ and each element of ${\bf H}_k \in \mathbb{F}_p^{d\times m}$ is drawn independently and uniformly from $\mathbb{F}_p$ for {\em some} $d$ and $m = \gamma d$. Note that the space expansion factor $\gamma$ is a constant and we allow scaling of the spatial dimension $d$ and $m$ while retaining their ratio $\gamma = m/d$. The spatial normalized capacity is the supremum of the achievable rate, $\overline{C}(\gamma) = \sup_d \overline{R}(\gamma) = \sup_d R/d$.

Next we explain why we allow spatial normalization, in spite of the fact that symbol extension along the key block domain already appears in the rate definition (refer to (\ref{rate})). The reason is that key block normalization creates {\em structured} (specifically, block diagonal with the same block) key matrices, which are more challenging to deal with, while spatial normalization creates fully generic key matrices. An example might help to illustrate this point. Suppose $N=2, E=2$ and $\gamma = 2$, i.e., each receiver sees a generic key subspace that has half dimension of the overall key space. When $d = 1$ and $m = \gamma d = 2$, for one block each key $z_k = {\bf H}_k {\bf s}$ is a generic linear combination of $2$ basis symbols and ${\bf H}_k \in \mathbb{F}_p^{1\times 2}$. Now consider spatial scaling by $d = 3$ and key block scaling by $L = 3$.
\begin{eqnarray}
\mbox{Spatial Extension:} && z_k =  \underbrace{ {\bf H}_k^{\mbox{\scriptsize ext}}}_{3 \times 6}  \underbrace{\bf s}_{6\times 1}, 
~\mbox{where each element of ${\bf H}_k^{\mbox{\scriptsize ext}}$ is randomly drawn}; \\
\mbox{Key Block Extension:} && Z_k = 
\left[ \begin{array}{c}
Z_k(1) \\
Z_k(2) \\
Z_k(3)
\end{array}
\right]  = 
\left[ \begin{array}{ccc}
{\bf H}_k & {\bf 0} & {\bf 0}\\
 {\bf 0} & {\bf H}_k & {\bf 0}\\
{\bf 0} & {\bf 0} & {\bf H}_k
\end{array}
\right]_{3\times 6}
\left[ \begin{array}{c}
{\bf s}(1) \\
{\bf s}(2) \\
{\bf s}(3)
\end{array}
\right]_{6\times 1}
\end{eqnarray}
where in key block extension, each block has identical distributions so that the linear combining coefficients must remain the same. Because of the constant diagonal key matrix structure, the generic secure groupcast problem is challenging without spatial normalization when we need symbol extensions (e.g., when the rate is non-integer) but the keys are no longer generic. In wireless parlance, key block extension corresponds to constant wireless channels with limited {\em diversity} \cite{Bresler_Tse_Diversity, Li_Ozgur, Hong_Caire}, which is also a challenging problem in degrees of freedom (DoF) studies and spatial normalization is exactly the remedy and a commonly used metric in wireless literature \cite{Wang_Gou_Jafar_subspace, Wang_Sun_Jafar, Sridharan_Yu, Liu_Yang}. 
Therefore, motivated by literature on DoF in wireless communications, we bring spatial normalization to generic secure groupcast and focus on the spatial normalized capacity for the setting with $N=2, E=2$. We note that the capacity when spatial normalization is not allowed, i.e., the capacity $C$ as a function of constant $d, m$, may not be equal to $\overline{C}(\gamma) d$ for every $d$ and remains an open problem in general.

We are now ready to present our results on the spatial normalized capacity for generic secure groupcast with $N=2, E=2$, in the following theorem.

\begin{figure}[h]
\begin{center}
\includegraphics[width=3 in]{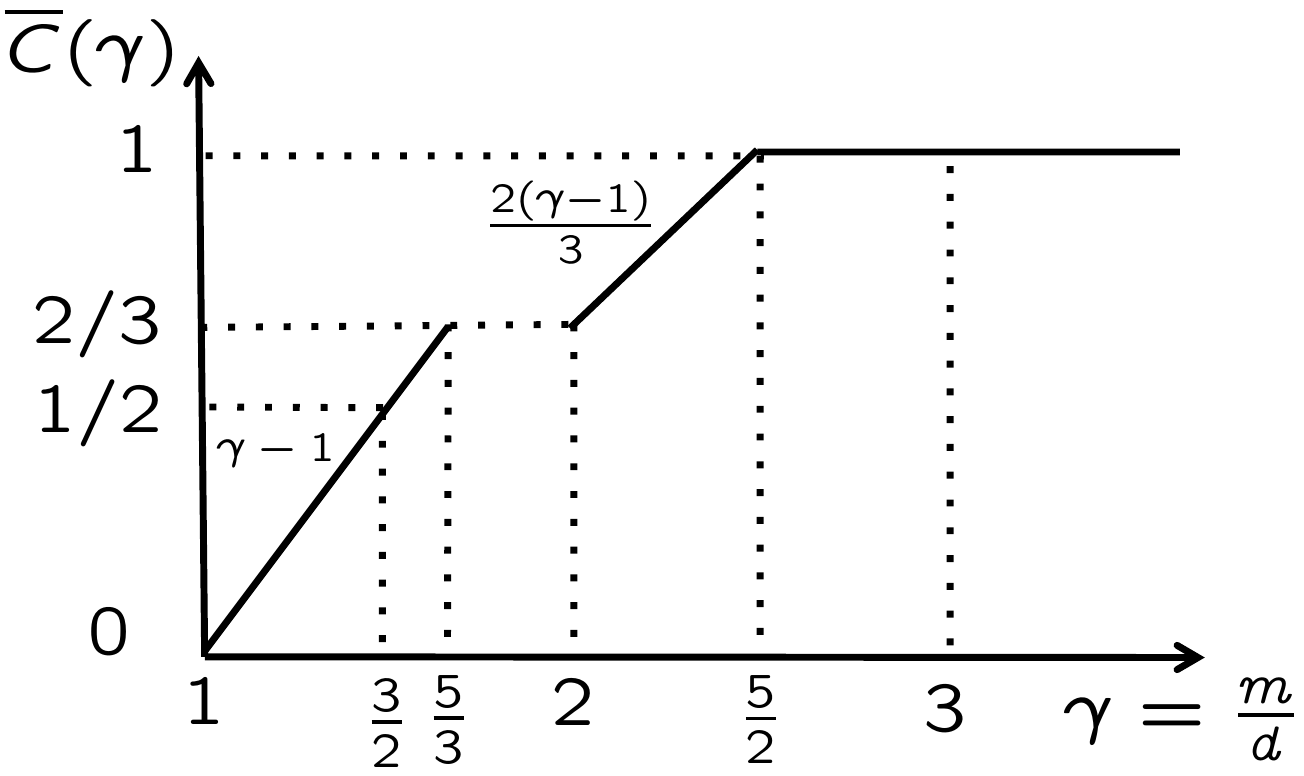}
\caption{\small For generic secure groupcast with $N=2, E= 2$, the spatial normalized capacity $\overline{C}(\gamma)$ is characterized except when $5/3 < \gamma < 2$.}
\label{fig:22}
\end{center}
\end{figure}

\begin{theorem}\label{thm:22}
For generic secure groupcast with $2$ qualified receivers and $2$ eavesdropping receivers, the spatial normalized capacity is
\begin{eqnarray}
\overline{C}(\gamma) = \left\{
\begin{array}{ll}
1, & \gamma \geq 5/2 \\
2(\gamma-1)/3, & 2 \leq \gamma \leq 5/2\\
\gamma - 1, & 1 \leq \gamma \leq 5/3
\end{array}
\right.
\end{eqnarray}
almost surely.
\end{theorem}

The result in Theorem \ref{thm:22} is plotted in Fig.~\ref{fig:22}. Compared with Theorem \ref{thm:generic}, the small $\gamma$ regime where $\overline{C}(\gamma) = \gamma - 1$ is extended from $1\leq \gamma \leq 3/2$ to include $3/2 \leq \gamma \leq 5/3$ and the large $\gamma$ regime where $\overline{C}(\gamma) = 1$ is extended from $\gamma \geq 3$ to include $5/2 \leq \gamma \leq 3$. In addition, a new regime  $2 \leq \gamma \leq 5/2$ is established, where $\overline{C}(\gamma)  = 2(\gamma-1)/3$ and a converse that is tighter than the conditional entropy bound (\ref{con:rate}) in Theorem \ref{thm:con} is required. The remaining regime where $5/3\leq \gamma \leq 2$ is open. The proof of Theorem \ref{thm:22} is deferred to Section \ref{sec:22} and we give an example of $\gamma = 2$ here.

\begin{example}\label{ex:22}
($\gamma = 2$)
We show that when $\gamma = 2$, $\overline{C}(\gamma) = 2/3$. An intuitive explanation of the converse result for linear schemes is as follows. Suppose rate $R$ is achievable. As $\gamma = 2$ so that the $N=2$ qualified receivers have independent keys almost surely, then the transmit signal size must be at least $L_X \geq 2R$ (refer to (\ref{con:beta})), i.e., the dimension of the key space of $X$ must be at least $2R$, $\mbox{rank}({\bf V}) \geq 2R$ (refer to (\ref{eq:linear})). Each eavesdropping key space has dimension $d$ ($\mbox{rank}({\bf H}_k) = d$), so that its overlap with the key space of $X$ is at least $d + L_X - m$, i.e., $\mbox{rank}({\bf U}_e) \geq d+L_X-m, e \in \{3,4\}$ (see (\ref{eq:proj})). From the security constraint (\ref{eq:sec}), the projection of the message in $X$ to ${\bf U}_e$ must be zero, i.e., $[{\bf U}_3; {\bf U}_4] {\bf V}_W = {\bf 0}$. Except from the projection to ${\bf U}_3, {\bf U}_4$, the orthogonal space of the message in $X$ has dimension at most $L_X - 2 (d + L_X - m)$, from which the message can be recovered by the correctness constraint (\ref{eq:corr}). Thus 
\begin{eqnarray}
 L_X - 2(d + L_X -m) &\geq& R \\
\Rightarrow ~~~~~~~~~~~~~~~~~~~~2m - 2d &\geq& R + L_X \geq 3R\\
\Rightarrow ~~~~~~~~~~~~~~~~ \overline{R}(\gamma) = R/d &\leq& 2(\gamma -1)/3 \overset{\gamma=2}{=} 2/3. \label{eq:22in}
\end{eqnarray}
It is not hard to translate the above argument to an information theoretic converse. In fact, we have further generalized it to cover a larger range of parameters (see Theorem \ref{thm:newcon}).

Interestingly, the converse argument above naturally leads us to the optimal achievable scheme. To this end, consider $d = 3, m = \gamma d = 6$ (note that for spatial normalized rate, we may pick $d, m$ values as long as the ratio $r = m/d$ is what we want). We present a coding scheme that achieves $R = 2d/3 = 2$. Following the insights from (\ref{eq:22in}), the transmit signal shall have dimension $2R = 4$ and does not need much special structure, thus it is set as 
\begin{eqnarray}
&& X = {\bf V}_W W + {\bf V} {\bf s}, {\bf V}_W \in \mathbb{F}_p^{4\times 2}, W \in \mathbb{F}_p^{2\times 1}, {\bf V} \in \mathbb{F}_p^{4\times 6}, {\bf s} \in \mathbb{F}_p^{6\times 1}, \notag \\
&&\mbox{where}~{\bf V} = [{\bf H}_1(1:2,:); {\bf H}_2(1:2,:)]~\mbox{and ${\bf V}_W$ will be specified later.} \label{eq:s22}
\end{eqnarray}
That is, the first two rows of the key in $X$ are from the key known to qualified Receiver $1$ and the last two key rows in $X$ are known to qualified Receiver $2$.
Next we identify the overlap of the key space of $X$ (i.e., $\mbox{rowspan}({\bf V})$) and the key space of each eavesdropping Receiver $e \in \{3,4\}$, i.e., $\mbox{rowspan}({\bf H}_e), {\bf H}_e \in \mathbb{F}_p^{3\times 6}$. For $e \in \{3,4\}$,
\begin{eqnarray}
&& \mbox{rowspan}({\bf U}_e {\bf V}) = \mbox{rowspan}({\bf P}_e {\bf H}_e) = \mbox{rowspan}({\bf V}) \cap \mbox{rowspan}({\bf H}_e), ~{\bf U}_e \in \mathbb{F}_p^{1 \times 4}, {\bf P}_e \in \mathbb{F}_p^{1\times 3} \notag\\
&\Rightarrow& 
\left[{\bf U}_e ~-{\bf P}_e\right]_{1 \times 7} \left[\begin{array}{cc}
{\bf V} \\
{\bf H}_e
\end{array}
\right]_{7 \times 6} = {\bf 0}_{1\times 6}, ~\mbox{i.e., ${\bf U}_e$ can be obtained from the left null space.}
\end{eqnarray}
Note that matrices ${\bf V}$ and ${\bf H}_e$ are generic so that the left null space has $1$ dimension with high probability and ${\bf U}_e$ exists. 
We are now ready to specify ${\bf V}_W$. From the security constraint (\ref{eq:sec}), we have
\begin{eqnarray}
\left[\begin{array}{cc}
{\bf U}_3 \\
{\bf U}_4
\end{array}
\right]_{2 \times 4} {\bf V}_W= {\bf 0}_{2\times 2}, ~\mbox{i.e., ${\bf V}_W$ can be set as column vectors from the right null space.} \label{eq:vw}
\end{eqnarray}
Finally, to guarantee correctness (\ref{eq:corr}), we need to ensure that qualified Receiver $1$ can obtain $W$ from the first two rows of $X$ (as the keys from the first two rows are known) and the qualified Receiver $2$ can obtain $W$ from the last two rows,
\begin{eqnarray}
&& \mbox{rank} \big( {\bf V}_W(1:2, :) \big) = 2, ~~\mbox{rank} \big( {\bf V}_W(3:4, :) \big) = 2.
\end{eqnarray}
The assignment of ${\bf V}_W$ from (\ref{eq:vw}) satisfies the above two rank constraints almost surely, because the key spaces of each receiver are generic (the coefficients of ${\bf H}_k$ are randomly drawn). The detailed proof of this observation is based on the Schwartz-Zippel lemma and appears in Section \ref{sec:22}.
\end{example}

The converse required for Theorem \ref{thm:22} is a special case of the following general converse result, whose idea generalizes that described in Example \ref{ex:22}.
\begin{theorem}\label{thm:newcon}
For the secure groupcast problem to the first $N$ of $K$ receivers, suppose 
the keys of the eavesdropping receivers are independent, i.e.,
\begin{eqnarray}
H(z_{N+1:K}) = \sum_{e = N+1}^K H(z_e), \label{eq:newind}
\end{eqnarray}  
then we have 
\begin{eqnarray}
R + (K-N-1) I(X; W, Z_{1:K} | U_{\mathcal{E}})/L \leq (K-N) H(z_{1:K} | u_{\mathcal{E}}) - \sum_{e = N+1}^K H(z_e | u_{\mathcal{E}})
\end{eqnarray}
where $u_\mathcal{E}$ satisfies $H(u_\mathcal{E} | z_e) = 0, \forall e \in [N+1:K]$, i.e., $u_\mathcal{E}$ is known to all eavesdropping receivers. 
\end{theorem}

The proof of Theorem \ref{thm:newcon} is presented in Section \ref{sec:newcon}.

\subsection{$N>2, E>2$ and Interference Alignment}
In this section, we fix $\gamma = 2$ and add more qualified or eavesdropping receivers to the $N=2, E=2$ generic secure groupcast system to see if the capacity $\overline{C}(\gamma = 2) = 2/3$ will change. Surprisingly, through various forms of interference alignment, including additional receivers may not hurt.

\begin{example}\label{ex:n3}
(Increasing $N$) Before including additional qualified receivers into the generic secure groupcast system with $N=2, E=2$, let us first review the insights for the achievable scheme with the optimal rate $\overline{R}(\gamma = 2) = 2/3$. From Example \ref{ex:22}, we set $d = 3$ so that $m = \gamma d = 6$, i.e., each receiver has $3$ generic linear combinations of $6$ basis systems ${\bf s} = s_{1:6}$ as the key. To achieve rate $R = 2d/3 = 2$, i.e., send $L_W = 2$ message symbols over $L=1$ key block, the transmit signal $X$ has $L_X = 4$ symbols (refer to (\ref{eq:s22})). The essential components are as follows.
\begin{enumerate}
\item The $4$-dimensional key space of $X$ consists of {\em random} $2$ dimensions from qualified key $z_1$ and {\em random} $2$ dimensions from qualified key $z_2$, so that each qualified receiver can decode $2$ equations on message symbols. 
\item The $4$-dimensional key space of $X$ has $1$-dimensional overlap each with eavesdropping key $z_3$ and $z_4$, respectively. To ensure security, the $2$-dimensional message space of $X$ is set to be orthogonal to the $2$-dimensional overlap (both overlaps), which exists as $X$ has $4$ dimensions.
\end{enumerate}
Now if we increase $N$ from $2$ to $3$ (i.e., receivers $1$ to $3$ are qualified and receivers $4, 5$ are eavesdropping) and wish to achieve the same rate $R = 2$, then we need to ensure that the key space of $X$ has a $2$-dimensional overlap with that of the additional qualified receiver. To this end, we can no longer pick $2$ random dimensions from $z_1$ and $z_2$ each, as such a random $4$-dimensional key space will overlap with the newly added $3$-dimensional qualified key $z_3$ in $3 + 4 - 6 = 1$ dimension. Therefore, we need to pick a $4$-dimensional key space that has $2$-dimensional overlaps with each of $z_1, z_2, z_3$. As $\gamma = 2$, each $z_i, z_j$ pair has no overlap almost surely, so what we need to do is to pick a $2$-dimensional subspace of $z_3$ that aligns into the span of the direct sum of $2$-dimensional subspaces of $z_1, z_2$.
In other words, we only need to change the first point (on correctness) above while the second point (on security) can be treated similarly (as the subspace of $z_3$ is aligned into those of $z_1, z_2$ so that $z_3$ essentially does not appear).

The detailed scheme with $N=3$ is as follows. We first perform a change of basis operation so that $z_1$ and $z_2$ are symbols along the standard basis after the transformation.
\begin{eqnarray}
&& {\bf H}_{12} \triangleq [{\bf H}_1 ; ~{\bf H}_2]_{6\times 6} ~\mbox{has full rank almost surely},\\
&& \mbox{new basis}~\overline{\bf s}_{6\times 1} \triangleq {\bf H}_{12} {\bf s}, ~\mbox{i.e.,}~{\bf s} = {\bf H}_{12}^{-1} \overline{\bf s} \label{eq:ba}\\
&\Rightarrow& z_1 = {\bf H}_1 {\bf s} = {\bf H}_1 {\bf H}_{12}^{-1} \overline{\bf s} = [{\bf I}_{3\times 3} ~{\bf 0}_{3\times 3}] \overline{\bf s} = \overline{s}_{1:3}\\
&& z_2 ={\bf H}_2 {\bf s} = {\bf H}_2 {\bf H}_{12}^{-1} \overline{\bf s} = [{\bf 0}_{3\times 3}~{\bf I}_{3\times 3}] \overline{\bf s} = \overline{s}_{4:6}\\
&& z_k = {\bf H}_k {\bf s} = \underbrace{ {\bf H}_k {\bf H}_{12}^{-1} }_{\triangleq [\overline{\bf H}_k^{[1]}~ \overline{\bf H}_k^{[2]}]} \overline{\bf s} =  \underbrace{\overline{\bf H}_k^{[1]}}_{3\times 3} \overline{s}_{1:3} + \underbrace{ \overline{\bf H}_k^{[2]}}_{3\times 3} \overline{s}_{4:6}, ~\forall k \in \{3,4,5\}.
\end{eqnarray}
The transmit signal is set as
\begin{eqnarray}
&& X = {\bf V}_W W + {\bf V} \overline{s}_{1:6} = {\bf V}_W W + 
\left[ \begin{array}{cc}
{\bf V}_1  & {\bf 0}_{2\times 3}\\
{\bf 0}_{2\times 3} & {\bf V}_2 \\
\end{array} 
\right] \overline{s}_{1:6}
=  {\bf V}_W W + 
\left[ \begin{array}{c}
{\bf V}_1 \overline{s}_{1:3} \\
{\bf V}_2 \overline{s}_{4:6} \\
\end{array} 
\right] \label{eq:vx}
\end{eqnarray}
where $X \in \mathbb{F}_p^{4\times 1}, {\bf V}_W \in \mathbb{F}_p^{4\times 2}, W \in \mathbb{F}_p^{2\times 1}, {\bf V} \in \mathbb{F}_p^{4 \times 6}, 
{\bf V}_1, {\bf V}_2 \in \mathbb{F}_p^{2 \times 3}$. Through the above design of the key space of $X$, i.e., $\mbox{rowspan}({\bf V})$, $z_1$ and $z_2$ each has a $2$-dimensional overlap, i.e., $\mbox{rowspan}({\bf V}_1)$ and $\mbox{rowspan}({\bf V}_2)$, and we consider $z_3$. We wish to guarantee the existence of a $2\times 3$ matrix ${\bf V}_3$ so that
\begin{eqnarray}
&& {\bf V}_3 z_3 = {\bf V}_3 (\overline{\bf H}_3^{[1]} \overline{s}_{1:3} + \overline{\bf H}_3^{[2]} \overline{s}_{4:6}) ~\mbox{can be obtained from}~{\bf V}_1 \overline{s}_{1:3}, {\bf V}_2 \overline{s}_{4:6} \\
&\Leftarrow& {\bf V}_1 = {\bf V}_3 \overline{\bf H}_3^{[1]}, {\bf V}_2 = {\bf V}_3 \overline{\bf H}_3^{[2]} \label{eq:v1}\\
&\Leftarrow& \mbox{We generate}~{\bf V}_3~\mbox{generically and set}~{\bf V}_1, {\bf V}_2~\mbox{following (\ref{eq:v1})}.
\end{eqnarray}
The alignment constraints in (\ref{eq:v1}) and the solution are similar to those in $2$ user wireless $X$ network \cite{Jafar_Shamai}. 
After ${\bf V}$ is specified, ${\bf V}_W$ is generated in the same manner as the $N=2$ case (see (\ref{eq:vw}) for detailed steps),
\begin{eqnarray}
\mbox{rowspan}({\bf V}_W) \perp \mbox{rowspan}({\bf V}) \cap \mbox{rowspan}(\overline{\bf H}_4),  \mbox{rowspan}({\bf V}) \cap \mbox{rowspan}(\overline{\bf H}_5) \label{eq:vw1}
\end{eqnarray}
so that security (\ref{eq:sec}) is guaranteed. Note that ${\bf V}$ is fully determined by qualified keys $z_1, z_2, z_3$.
For correctness (\ref{eq:corr}), we require that ${\bf V}_W(1:2, :), {\bf V}_W(3:4, :), {\bf V}_W(1:2, :)+{\bf V}_W(3:4, :)$ each has rank $2$ (which holds almost surely by showing each determinant polynomial is not the zero polynomial) so that Receiver $1$, $2$, and $3$ can decode the desired message, respectively.

Next, what if $N$ is further increased to $4$? We use the same transmit signal structure (\ref{eq:vx}) and naturally have additional alignment constraints due to the new qualified Receiver $4$, i.e., we wish to guarantee that both $z_3$ and $z_4$ have a $2$-dimensional overlap with ${\bf V}\overline{s}_{1:6}$. This is realized by setting $2\times 3$ matrices ${\bf V}_3, {\bf V}_4$ so that
\begin{eqnarray}
&& {\bf V}_3 z_3 = {\bf V}_3 (\overline{\bf H}_3^{[1]} \overline{s}_{1:3} + \overline{\bf H}_3^{[2]} \overline{s}_{4:6})  ~\mbox{can be obtained from}~{\bf V}_1 \overline{s}_{1:3}, {\bf V}_2 \overline{s}_{4:6} \\
&& {\bf V}_4 z_4 = {\bf V}_4 (\overline{\bf H}_4^{[1]} \overline{s}_{1:3} + \overline{\bf H}_4^{[2]} \overline{s}_{4:6}) ~\mbox{can be obtained from}~{\bf V}_1 \overline{s}_{1:3}, {\bf V}_2 \overline{s}_{4:6} \\
&\Leftarrow& \mbox{rowspan}({\bf V}_1) = \mbox{rowspan}({\bf V}_3 \overline{\bf H}_3^{[1]}) = \mbox{rowspan}({\bf V}_4 \overline{\bf H}_4^{[1]}) \notag \\
&&  \mbox{rowspan}({\bf V}_2) = \mbox{rowspan}({\bf V}_3 \overline{\bf H}_3^{[2]}) = \mbox{rowspan}({\bf V}_4 \overline{\bf H}_4^{[2]} ) \label{eq:v2} \\
&\Leftarrow& \mbox{rowspan}({\bf V}_3) = \mbox{rowspan}({\bf V}_3 \underbrace{ \overline{\bf H}_3^{[2]} (\overline{\bf H}_4^{[2]})^{-1} \overline{\bf H}_4^{[1]}  (\overline{\bf H}_3^{[1]})^{-1} }_{\triangleq \overline{\bf H}_c} )\\
&\Leftarrow& \mbox{We set}~{\bf V}_3^T~\mbox{as the eigenvectors of $\overline{\bf H}_c^T$ and then set}~{\bf V}_1, {\bf V}_2, {\bf V}_4~\mbox{following (\ref{eq:v2})}.
\end{eqnarray}
The alignment constraints in (\ref{eq:v2}) and the solution are similar to those in $3$ user interference network \cite{Cadambe_Jafar_int}. 
The assignment of ${\bf V}_W$ is exactly the same as above (see (\ref{eq:vw1})). As the overlapping key space of $X$ for the additional qualified receiver is aligned into those for the original qualified receivers $1, 2$, the proof of correctness and security remains the same.

Finally, suppose we have a large integer $N$. If we follow the same idea above, the linear systems will be over-constrained. Following (\ref{eq:v2}), we need to find ${\bf V}_q, q \in [1:N]$ so that
\begin{eqnarray}
&& \mbox{rowspan}({\bf V}_1) \approx \mbox{rowspan}({\bf V}_3 \overline{\bf H}_3^{[1]}) \approx \mbox{rowspan}({\bf V}_4 \overline{\bf H}_4^{[1]}) \approx \cdots \approx \mbox{rowspan}({\bf V}_N \overline{\bf H}_N^{[1]}) \notag\\
&&  \mbox{rowspan}({\bf V}_2) \approx \mbox{rowspan}({\bf V}_3 \overline{\bf H}_3^{[2]}) \approx \mbox{rowspan}({\bf V}_4 \overline{\bf H}_4^{[2]} ) \approx \cdots \approx \mbox{rowspan}({\bf V}_N \overline{\bf H}_N^{[2]}). \label{eq:vv}
\end{eqnarray} 
Such over-constrained linear systems for large $N$ are a canonical challenge in interference alignment. 
Exact solutions may not exist and we have replaced the exact equality `$=$' with approximate equality `$\approx$'.
A well-known technique is to employ CJ asymptotic interference alignment \cite{Cadambe_Jafar_int}, which however, requires {\em diagonal} channel (key) matrices. The solution of (\ref{eq:vv}) turns out to be the generalization of CJ asymptotic interference alignment from single antenna to multiple antenna wireless systems \cite{Gou_Jafar_MIMO, Sun_Gou_Jafar, Maddah_miso}.
In short, when $\overline{\bf H}_q^{[1]}, \overline{\bf H}_q^{[2]}, q \in [3:N]$ are generic diagonal matrices, we can find an asymptotic interference alignment based solution to (\ref{eq:vv}). The details are deferred to the proof of the theorem stated below in Section \ref{sec:nbig}.
\end{example}

\begin{theorem}\label{thm:nbig}
For generic secure groupcast with either $N > 2, E= 2$ or $N=2, E>2$, if $\gamma = 2$ and the keys $z_k, k \in [1:N+E]$ are 
\begin{eqnarray}
z_k = {\bf H}_k^{[1]} {s}_{1:d} + {\bf H}_k^{[2]} {s}_{d+1:2d}
\end{eqnarray}
where ${\bf H}_k^{[1]}, {\bf H}_k^{[2]} \in \mathbb{F}_p^{d\times d}$ are generic diagonal matrices, i.e., each diagonal element of ${\bf H}_k^{[1]}, {\bf H}_k^{[2]}$ is drawn independently and uniformly from $\mathbb{F}_p$ and all non-diagonal elements are equal to zero, then 
$\overline{C}(\gamma = 2) = 2/3$. When $E > 2$, the achievable scheme has $\epsilon$ leakage, i.e., $I(W; X, Z_k) = o(d)$.
\end{theorem}

\begin{remark}\label{remark:dia}
Note that Theorem \ref{thm:nbig} requires generic diagonal key matrices, which satisfy commutativity - the key for alignment to be possible in over-constrained systems. The case with generic full key matrices (each element randomly drawn with no fixed zeros, see (\ref{eq:generic})) is generally open.
\end{remark}

Note that Theorem \ref{thm:nbig} states that we may either increase $N$ or $E$, without decreasing the capacity. Interestingly, the case of increasing $E$ turns out to be somewhat the dual of increasing $N$. 
When only $N$ is increased, we design the key space of $X$, $\mbox{rowspan}({\bf V})$ first (fully determined by qualified keys $z_1, \cdots, z_N$), and then find the overlaps with the eavesdropping keys $z_{N+1}, \cdots, z_K$ to determine the message space of $X$, $\mbox{rowspan}({\bf V}_W)$ (refer to (\ref{eq:vw1})), i.e., from ${\bf V}$ to ${\bf V}_W$. 
When only $E$ is increased, the order is reversed. Specifically, we design the overlap of the key space of $X$ with the eavesdropping keys $z_{N+1}, \cdots, z_K$ first (fully determined by alignment constraints among the eavesdropping keys) such that $\mbox{rowspan}({\bf V}_W)$ is set as the orthogonal space, and then find the key space of $X$, $\mbox{rowspan}({\bf V})$ to have the desired determined overlaps with the eavesdropping keys, i.e., from ${\bf V}_W$ to ${\bf V}$. Further, $\mbox{rowspan}({\bf V})$ must be designed so that its overlap with each qualified key has sufficient dimensions to ensure correctness. The details are presented in Section \ref{sec:nbig}.

\begin{remark}\label{remark:ne}
We have only considered the case of either increasing $N$ or increasing $E$ above. What if we simultaneously increase $N$ and $E$? This is an open problem and the ideas presented above may not suffice because
the design of ${\bf V}$ (fully determined by qualified keys for large $N$) and ${\bf V}_W$ (fully determined by eavesdropping keys for large $E$) is not compatible in general.
\end{remark}

\section{Proofs}
\subsection{Proof of Lemma \ref{lemma:sp4}: Capacity of $\mbox{SG}_4$}\label{sec:sp4}
We present the achievability and converse proofs in the following two sections.
\subsubsection{Achievability: $R \geq 4$}
To send $L_W = 4$ message symbols $W_{1:4}$ with $L=1$ key block, we set the transmit signal $X = (X_1; X_2)$ as follows. Note that $X$ contains two independent parts $X_1 \in \mathbb{F}_p^{4\times 1}$ and $X_2 \in \mathbb{F}_p^{4\times 1}$.
\begin{eqnarray}
X_1 = \left( \begin{array}{c}
W_1 + s_4 \\
W_2 + s_5 + s_6 \\
-2W_1 - 3W_2 + s_8 \\
W_1 + 2W_2 + s_7 + s_9 \\
\end{array}
\right), ~~
X_2 = \left( \begin{array}{c}
W_3 + s_{10} \\
W_4 + s_{11} + s_{12} \\
-2W_3 - 3W_4 + s_{14} \\
W_3 + 2W_4 + s_{13} + s_{15} \\
\end{array}
\right).
\end{eqnarray}

Correctness (\ref{eq:corr}) follows from the observation that qualified Receiver $1$ knows $s_{4:6}, s_{10:12}$ and can obtain the message symbols $W_{1:4}$ from the first two rows of $X_1, X_2$, and qualified Receiver $2$ knows $s_{7:9}, s_{13:15}$ and can obtain the message symbols $W_{1:4}$ from the last two rows of $X_1, X_2$. 

Consider security (\ref{eq:sec}) and the eavesdropping Receiver $3$. Note that the overlap of the key space of $X_1$ and the eavesdropping key $z_3$ is 
\begin{eqnarray} 
s_4 + s_5 + s_6 + s_8 + s_7 + s_9,
\end{eqnarray}
i.e., the sum of all four rows of $X_1$. The projection of the message symbols to this space is
\begin{eqnarray}
W_1 + W_2 + (-2W_1-3W_2) + (W_1 + 2W_2) = 0
\end{eqnarray}
so that nothing is revealed. The case with $X_2$ is similar, i.e., from $X_1, X_2, z_3$, we obtain no information about $W$. Consider eavesdropping Receiver $4$. The overlap of the key space of $X_1$ and the eavesdropping key $z_4$ is
\begin{eqnarray} 
2s_4 + (s_5 + s_6) + 3s_8 + 4(s_7 + s_9) = 2(s_4 + 2s_7) + (s_5 + 3s_8) + (s_6+4s_9),
\end{eqnarray}
i.e., $[2, 1, 3, 4]_{1 \times 4} \times X_1$. The projection of the message symbols to this space is
\begin{eqnarray}
2 W_1 +  W_2 + 3 (-2W_1-3W_2) + 4 (W_1 + 2W_2) = 0
\end{eqnarray}
so that nothing is revealed from $(X_1, X_2, z_4)$. The achievability proof of Lemma \ref{lemma:sp4} is complete.

\subsubsection{Converse: $R \leq 4$}
We use Theorem \ref{thm:newcon} and Theorem \ref{thm:con}. Set $u_{\mathcal{E}} = s_{1:3}$ in Theorem \ref{thm:newcon}, then we have 
\begin{eqnarray}
 \Rightarrow R + I(X; W, Z_{1:4} | U_{\mathcal{E}})/L &\leq& 2 H(z_{1:4} | u_{\mathcal{E}}) - H(z_3 | u_{\mathcal{E}}) - H(z_4 | u_{\mathcal{E}}) \\
&=& 2 \times 12 - (6 + 6) = 12 .
\end{eqnarray}
Set $\mathcal{Q} = \{1,2\}$ and $u_{e} = u_{\mathcal{E}} = s_{1:3}$ in Theorem \ref{thm:con}, then we have
\begin{eqnarray}
I(X; W, Z_{1:4} | U_{\mathcal{E}})/L &\geq& 2R - \Big( H(z_3 | u_\mathcal{E}) + H(z_4 | u_\mathcal{E}) - H(z_3, z_4 | u_\mathcal{E}) \Big) \\
&=& 2R - (6 + 6 - 12) = 2R.
\end{eqnarray}
Combining with the above two inequalities, we have the desired outer bound,
\begin{eqnarray}
3R \leq 12 &\Rightarrow& R \leq 4 .
\end{eqnarray}

\subsection{Proof of Theorem \ref{thm:generic}: Extreme $\gamma$ Regimes}\label{sec:generic}
\subsubsection{Large $\gamma$: $\gamma \geq \min(N+1, E+1)$}
The converse proof follows immediately from Theorem \ref{thm:con}. From (\ref{con:rate}), $\forall q \in [1:N], \forall e \in [N+1:K]$ we have 
\begin{eqnarray}
&& R \leq H(z_q | z_e) \leq H(z_q) = \mbox{rank}({\bf H}_q) \leq d.
\end{eqnarray}
Note that $R/d \leq 1$ always holds, e.g., there is no probabilistic argument involved.

The achievability proof has two parts. 
First, we show that when $\gamma \geq N+1$, $R = d$ is achievable. Consider $L=1$ key block, and the message has $d$ symbols, $W \in \mathbb{F}_p^{d\times 1}$. We set
\begin{eqnarray}
X = \left( \begin{array}{c}
W + z_1\\
W + z_2\\
\vdots\\
W + z_N
\end{array}
\right) \in \mathbb{F}_{p}^{Nd\times 1}
\end{eqnarray}
where $z_q \in \mathbb{F}_p^{d \times 1}, q \in [1:N]$ and `$+$' represents element-wise addition. Correctness constraint (\ref{eq:corr}) is always satisfied because each qualified Receiver $q$ can use $z_q$ to obtain $W$ from $X$. For security (\ref{eq:sec}), we require that $\forall e \in [N+1:K]$
\begin{eqnarray}
&& (z_1, z_2, \cdots, z_N) ~\mbox{is independent of}~ z_e \\
&\Leftarrow& {\bf H}_{\mathcal{Q} \cup \{e\}} \triangleq [{\bf H}_1 ; ~{\bf H}_2; ~\cdots; ~{\bf H}_N; ~{\bf H}_e]_{(N+1)d \times m} ~\mbox{has full row rank}~\label{eq:f1}
\end{eqnarray}
which holds almost surely because $m = \gamma d \geq (N+1)d$ and each element of ${\bf H}_{\mathcal{Q} \cup \{e\}}$ is drawn independently and uniformly from $\mathbb{F}_p$. Consider the determinant of any $(N+1)d \times (N+1)d$ sub-matrix of ${\bf H}_{\mathcal{Q} \cup \{e\}}$ and view the determinant as a polynomial, whose variables are the elements of ${\bf H}_{\mathcal{Q} \cup \{e\}}$. This polynomial is not the zero polynomial so that by the Schwartz-Zippel lemma \cite{Demillo_Lipton, Schwartz, Zippel}, the probability that the determinant is not zero approaches 1 as the field size $p$ approaches infinity, i.e., (\ref{eq:f1}) holds almost surely. As (\ref{eq:f1}) holds with probability approaching $1$ for each $e \in [N+1:K]$, the probability that (\ref{eq:f1}) holds for all $e \in [N+1:K]$ also approaches $1$, i.e., the security constraint (\ref{eq:sec}) is satisfied almost surely.

Second, we show that when $\gamma \geq E+1$, $R = d$ is achievable. Set $L=1$ and
\begin{eqnarray}
X = {\bf V}_W W + {\bf s}
\end{eqnarray}
where $X, {\bf s} \in \mathbb{F}_p^{m \times 1}, {\bf V}_W \in \mathbb{F}_p^{m \times d}, W \in \mathbb{F}_p^{d\times 1}$ and ${\bf V}_W$ is chosen so that
\begin{eqnarray}
\underbrace{ \left[\begin{array}{c}
{\bf H}_{N+1} \\
{\bf H}_{N+2} \\
\vdots\\
{\bf H}_{K}
\end{array}
\right]_{Ed \times m}}_{\triangleq {\bf H}_{\mathcal{E}}} {\bf V}_W = {\bf 0}_{Ed \times d}.
\end{eqnarray}
Note that $m = \gamma d \geq (E+1) d$, i.e., $m - Ed \geq d$, so the right null space of ${\bf H}_\mathcal{E}$ has at least $d$ dimensions, i.e., ${\bf V}_W$ exists and can be chosen as any $d$ linearly independent column vectors from the right null space. This choice of ${\bf V}_W$ ensures security (\ref{eq:sec}) and we verify correctness (\ref{eq:corr}). We require that $\forall q \in [1:N]$, from ${\bf H}_q X 
= {\bf H}_q {\bf V}_W W + z_q$,
we can decode $W$, i.e.,
\begin{eqnarray}
\mbox{rank}({\bf H}_q {\bf V}_W) = d. \label{eq:f2}
\end{eqnarray}
We similarly invoke the Schwartz-Zippel lemma \cite{Demillo_Lipton, Schwartz, Zippel}. To this end, view the determinant of ${\bf H}_q {\bf V}_W$ as a polynomial in variables of the elements of ${\bf H}_k, k \in [1:K]$. This polynomial is not the zero polynomial because it is not always zero, e.g., we may assign the matrices ${\bf H}_k$ as follows,
\begin{eqnarray}
{\bf H}_{\mathcal{E}} = \left[\begin{array}{ccc}
{\bf I}_{Ed\times Ed} & {\bf 0}_{Ed \times (m - Ed)}
\end{array}
\right], ~{\bf H}_q = \left[\begin{array}{ccc}
{\bf 0}_{d\times (m-d)} & {\bf I}_{d \times d}
\end{array}
\right]
\end{eqnarray}
so that
\begin{eqnarray}
{\bf V}_W = \left[
\begin{array}{c}
{\bf 0}_{(m-d)\times d}\\
{\bf I}_{d \times d}
\end{array}
\right], ~~ {\bf H}_q {\bf V}_W = {\bf I}_{d\times d} ~\Rightarrow~ \det({\bf H}_q {\bf V}_W) = 1 \neq 0.
\end{eqnarray}
As a result, the non-zero polynomial will not be zero with probability approaching $1$ as $p \rightarrow \infty$. As (\ref{eq:f2}) holds with probability approaching $1$ for each $q \in [1:N]$, the probability that (\ref{eq:f2}) holds for all $q \in [1:N]$ also approaches $1$, i.e., the correctness constraint (\ref{eq:corr}) is satisfied almost surely.

\subsubsection{Small $\gamma$: $\gamma \leq \max(1+1/N, 1+1/E)$}
We use Theorem \ref{thm:con} to prove the converse. From (\ref{con:rate}), $\forall q \in [1:N], \forall e \in [N+1:e]$ we have
\begin{eqnarray}
R \leq H(z_q | z_e) = H(z_q, z_e) - H(z_e) \leq H({\bf s}) - H(z_e) = m - \mbox{rank}({\bf H}_e)
\end{eqnarray}
which is equal to $m - d$ almost surely, as ${\bf H}_e$ contains $d$ generic rows. Thus $R/d \leq m/d - 1 = \gamma - 1$ almost surely.

We now consider the achievability proof, which has two parts. First, we show that, when $\gamma \leq 1+1/N$, $R = m - d$ is achievable. Set $L=1$ and
\begin{eqnarray}
X = W + {\bf H}_\mathcal{Q} {\bf s}
\end{eqnarray}
where $X, W \in \mathbb{F}_{p}^{(m-d) \times 1}, {\bf H}_\mathcal{Q} \in \mathbb{F}_p^{(m-d) \times m}, {\bf s} \in \mathbb{F}_p^{m\times 1}$ and ${\bf H}_\mathcal{Q}$ is chosen so that
\begin{eqnarray}
&& {\bf H}_\mathcal{Q} = {\bf P}_1 {\bf H}_1 = {\bf P}_2 {\bf H}_2 = \cdots = {\bf P}_N {\bf H}_N \\
&\Leftarrow&  {\bf 0}_{(m-d)\times (N-1)m} = 
\left[\begin{array}{ccccc}
{\bf P}_1 & {\bf P}_2 & \cdots & {\bf P}_N
\end{array}
\right]_{(m-d) \times Nd}
\underbrace{\left[\begin{array}{ccccc}
{\bf H}_1 & {\bf H}_1 & \cdots & {\bf H}_1\\
-{\bf H}_2 & {\bf 0}_{d\times m} & \cdots & {\bf 0}_{d\times m}\\
{\bf 0}_{d\times m} & -{\bf H}_3 & \cdots & {\bf 0}_{d\times m}\\
\vdots & \vdots & \ddots & \vdots \\
{\bf 0}_{d\times m} & {\bf 0}_{d\times m} & \cdots & -{\bf H}_{N}
\end{array}
\right]_{Nd \times (N-1)m}}_{\triangleq {\bf H}_\mathcal{B}}, \notag\\
&& {\bf H}_\mathcal{Q} = {\bf P}_1 {\bf H}_1. \label{eq:o1}
\end{eqnarray}
Note that $Nd - (N-1)m = m - N(\gamma - 1)d \geq m-d$, so the left null space of ${\bf H}_\mathcal{B}$ has at least $m-d$ dimensions, i.e., ${\bf P}_q, q \in [1:N]$ exists and can be chosen as any $m-d$ linearly independent row vectors from the left null space. Then ${\bf H}_{\mathcal{Q}}$ exists and this choice of ${\bf H}_\mathcal{Q}$ ensures correctness (\ref{eq:corr}), because any qualified Receiver $q \in [1:N]$ can obtain ${\bf H}_\mathcal{Q} {\bf s} = {\bf P}_q z_q$ and then extract $W$ from $X$. Next, consider security (\ref{eq:sec}). We require that $\forall e \in [N+1:K]$
\begin{eqnarray}
{\bf H}_{\mathcal{Q}\cup \{e\}} \triangleq [{\bf H}_\mathcal{Q}; ~{\bf H}_e]_{m\times m} ~\mbox{has full rank almost surely} \label{eq:o2}
\end{eqnarray}
which follows from the Schwartz-Zippel lemma {\cite{Demillo_Lipton, Schwartz, Zippel}} and the determinant polynomial of ${\bf H}_{\mathcal{Q}\cup \{e\}}$ is not the zero polynomial (easy to see as ${\bf H}_\mathcal{Q}$ only depends on the qualified key matrices ${\bf H}_q, q \in [1:N]$, which is independent of the eavesdropping key matrix ${\bf H}_e$). Thus the security constraint (\ref{eq:sec}) is satisfied almost surely over large fields, i.e., when $p \rightarrow \infty$.

Second, we show that when $\gamma \leq 1+1/E$, $R = m-d$ is achievable. Set $L = 1$ and
\begin{eqnarray}
X = \left( \begin{array}{c}
{\bf H}_\mathcal{E} \hspace{0.02in} {\bf s} \\
W + {\bf H}_{\mbox{\scriptsize}rand} \hspace{0.03in} {\bf s}
\end{array}
\right)
\end{eqnarray}
where $X \in \mathbb{F}_p^{2(m-d) \times 1}, {\bf H}_\mathcal{E}, {\bf H}_{\mbox{\scriptsize}rand} \in \mathbb{F}_p^{(m-d)\times m}, W \in \mathbb{F}_p^{(m-d)\times 1}, {\bf s} \in \mathbb{F}_p^{m\times 1}$ and ${\bf H}_\mathcal{E}, {\bf H}_{\mbox{\scriptsize}rand}$ are chosen as follows.
\begin{eqnarray}
&& \mbox{Each element of ${\bf H}_{\mbox{\scriptsize}rand}$ is drawn independently and uniformly from $\mathbb{F}_p$ and}\\
&& \mbox{rowspan}({\bf H}_\mathcal{E}) = \mbox{rowspan}({\bf H}_{N+1}) \cap \mbox{rowspan}({\bf H}_{N+2}) \cap \cdots \cap \mbox{rowspan}({\bf H}_{K}) \\
&\Leftarrow& {\bf H}_\mathcal{E} = {\bf P}_{N+1} {\bf H}_{N+1} = {\bf P}_{N+2} {\bf H}_{N+2} = \cdots = {\bf P}_K {\bf H}_K, ~{\bf P}_e \in \mathbb{F}_p^{(m-d) \times d}, e \in [N+1:K]
\end{eqnarray}
where ${\bf H}_\mathcal{E}$ can be solved in the same manner as (\ref{eq:o1}) because the overlap of the row spaces of ${\bf H}_e$ has sufficient dimensions, i.e., $Ed - (E-1)m = m - E(\gamma - 1)d \geq m-d$. To ensure correctness (\ref{eq:corr}), we require
\begin{eqnarray}
[{\bf H}_\mathcal{E}; ~{\bf H}_q]_{m\times m}~\mbox{has full rank almost surely}~, \forall q \in [1:N]
\end{eqnarray}
whose proof follows similarly from that of (\ref{eq:o2}). Then each qualified Receiver $q$ can recover ${\bf s}$ from ${\bf H}_\mathcal{E}  \hspace{0.02in} {\bf s}$ and the key $z_q = {\bf H}_q {\bf s}$, and obtain ${\bf H}_{\mbox{\scriptsize}rand} \hspace{0.03in} {\bf s}$ (so that $W$ is decoded with no error). To ensure security (\ref{eq:sec}), we require
\begin{eqnarray}
{\bf H}_{\mbox{\scriptsize} rand\cup\{e\}} \triangleq [{\bf H}_{\mbox{\scriptsize}rand}; ~{\bf H}_e]_{m\times m}~\mbox{has full rank almost surely}~, \forall e \in [N+1:K]
\end{eqnarray}
which follows from the Schwartz-Zippel lemma {\cite{Demillo_Lipton, Schwartz, Zippel}} and the determinant polynomial of ${\bf H}_{\mbox{\scriptsize} rand\cup\{e\}}$ is not the zero polynomial (trivial as we may find a realization of ${\bf H}_{\mbox{\scriptsize}rand}$ and ${\bf H}_k, k \in [1:K]$ such that ${\bf H}_{\mbox{\scriptsize} rand\cup\{e\}}$ is an identity matrix). To sum up, both correctness constraint (\ref{eq:corr}) and security constraint (\ref{eq:sec}) are satisfied almost surely over a sufficiently large field. The proof of Theorem \ref{thm:generic} is now complete.

\subsection{Proof of Theorem \ref{thm:22}: $N=E=2$} \label{sec:22}
As the regimes where $1 \leq \gamma \leq 3/2$ and $\gamma \geq 3$ have been covered by Theorem \ref{thm:generic}, we only need to consider the remaining three regimes, which are discussed sequentially as follows. 
\subsubsection{$5/2 \leq \gamma \leq 3$}
Converse follows from Theorem \ref{thm:con}. From (\ref{con:rate}), we have $R \leq H(z_q | z_e) \leq H(z_q) = \mbox{rank}({\bf H}_q) \leq d, \forall q \in \{1,2\}, \forall e \in\{3,4\}$.

Achievability of $R = d$ is similar to that of Example \ref{ex:22}. Set $L=1$ and the transmit signal as
\begin{eqnarray}
X =   {\bf V}_W W + {\bf V}{\bf s} = 
{\bf V}_W W + {\left[ \begin{array}{c}
{\bf H}_1 \\
{\bf H}_2
\end{array} \right]} {\bf s} =
{\bf V}_W W + \left[ \begin{array}{c}
z_1 \\
z_2
\end{array} \right]
\end{eqnarray}
where $X \in \mathbb{F}_p^{2d\times1}, {\bf V}_W \in \mathbb{F}_p^{2d\times d}, W \in \mathbb{F}_p^{d\times 1}, {\bf V} \in \mathbb{F}_{p}^{2d\times m}, {\bf s} \in \mathbb{F}_p^{m\times 1}, {\bf H}_1, {\bf H}_2 \in \mathbb{F}_p^{d\times m}$, and ${\bf V}_W$ is designed as follows,
\begin{eqnarray}
&& \underbrace{\left[\begin{array}{c}
{\bf U}_3 \\
{\bf U}_4
\end{array}
\right]}_{2(3d-m)\times 2d} {\bf V}_W = {\bf 0}, ~\mbox{i.e., ${\bf V}_W$ exists as $2d - 2(3d-m) \geq d~(\gamma \geq 5/2)$}, \label{eq:2p} \\
&& \mbox{where for $e \in \{3,4\}$,}~[{\bf U}_e~-{\bf P}_e]_{(3d-m) \times 3d}
\left[\begin{array}{c}
{\bf V}\\
{\bf H}_e
\end{array}
\right]_{3d\times m} = {\bf 0}.
\end{eqnarray}
Note that ${\bf V}$ is fully determined by ${\bf H}_1, {\bf H}_2$, which is independent of ${\bf H}_e$, so ${\bf U}_e$ (obtained from the overlap of the row space of ${\bf V}$ and the row space of ${\bf H}_e$) will have dimension $3d - m$ with high probability. Security (\ref{eq:sec}) is guaranteed by (\ref{eq:2p}). For correctness (\ref{eq:corr}), we require that
\begin{eqnarray}
{\bf V}_W(1:d, :), {\bf V}_W(d+1:2d, :)~\mbox{both have full rank}
\end{eqnarray}
which is proved by showing that the determinant polynomials (of variables from ${\bf H}_k, k \in [1:4]$) are not the zero polynomial so that by the Schwartz-Zippel lemma \cite{Demillo_Lipton, Schwartz, Zippel}, the two matrices have full rank almost surely as $p \rightarrow \infty$. The determinant polynomials are non-zero for the following realization of ${\bf H}_k$ so that they are not always zero. 
\begin{eqnarray}
&& {\bf H}_1 {\bf s} = s_{1:d}, ~{\bf H}_2 {\bf s} = s_{(d+1):2d} \\
&& {\bf H}_3 {\bf s} = (s_{1:(3d-m)}+s_{(d+1):(4d-m)}; s_{(2d+1):m}) \\
&& {\bf H}_4 {\bf s}= (s_{(3d-m+1):2(3d-m)}+s_{(4d-m+1):(7d-2m)}; s_{(2d+1):m}) \\
&& \left[\begin{array}{c}
{\bf U}_3\\
{\bf U}_4
\end{array}\right] = [
{\bf I}_{2(3d-m)\times 2(3d-m)}~{\bf 0}_{2(3d-m)\times(2m-5d)}~{\bf I}_{2(3d-m)\times 2(3d-m)}~{\bf 0}_{2(3d-m)\times(2m-5d)}
]\\
&&{\bf V}_W = 
[{\bf I}_{d \times d};
-{\bf I}_{d\times d}].
\end{eqnarray}
The proof of achievability when $5/2 \leq \gamma \leq 3$ is complete.

\subsubsection{$2\leq \gamma \leq 5/2$}
We first provide the converse proof. In Theorem \ref{thm:newcon}, we set $u_{\mathcal{E}} = ()$. Note that $\gamma \geq 2$ so the eavesdropping keys $z_3, z_4$ are independent almost surely, i.e., the condition of Theorem \ref{thm:newcon} is satisfied. Then we have
\begin{eqnarray}
R + I(X; W, Z_{1:4})/L \leq 2 H(z_{1:4}) - H(z_3) - H(z_4) = 2m - 2d
\end{eqnarray}
almost surely as the keys are generic.
Then we apply Theorem \ref{thm:con}. In (\ref{con:beta}), we set $\mathcal{Q} = \{1,2\}$ and it follows that
\begin{eqnarray}
I(X; W, Z_{1:4})/L  \geq 2R - ( H(z_1, z_2) - H(z_1) - H(z_2) ) = 2R
\end{eqnarray}
almost surely. Note that when $\gamma \geq 2$, the qualified keys $z_1, z_2$ are independent almost surely. 
Combining the two inequalities above, we have
\begin{eqnarray}
3R \leq 2m-2d  ~\Rightarrow~ \overline{R}(\gamma) = R/d \leq 2(m-d)/(3d) = 2(\gamma-1)/3
\end{eqnarray}
and the converse proof when $2\leq \gamma \leq 5/2$ is complete.

We next provide the achievability proof, which is very similar to that presented in the previous section and only the parameters need to be adjusted to match the current $\gamma$ regime. As the normalized rate $\overline{R}(\gamma) = 2(\gamma - 1)/3$ may not be an integer, we consider spatial extension by a factor of $3$, i.e., set $d' = 3d,  m' = 3m$ and show that $R = 2(m'- d')/3 = 2(m-d)$ is achievable when each key is $3d$ generic combinations of $3m$ basis symbols.
Set $L=1$ and 
\begin{eqnarray}
X =   {\bf V}_W W + {\bf V}{\bf s} = 
{\bf V}_W W + {\left[ \begin{array}{c}
{\bf H}_1(1:2(m-d), :) \\
{\bf H}_2(1:2(m-d), :)
\end{array} \right]} {\bf s} 
\end{eqnarray}
where $X \in \mathbb{F}_p^{4(m-d)\times1}, {\bf V}_W \in \mathbb{F}_p^{4(m-d)\times 2(m-d)}, W \in \mathbb{F}_p^{2(m-d)\times 1}, {\bf V} \in \mathbb{F}_{p}^{4(m-d)\times 3m}, {\bf s} \in \mathbb{F}_p^{3m\times 1}$. Note that $2(m-d) \leq 3d$ as $\gamma \leq 5/2$. ${\bf V}_W$ is designed so that
\begin{eqnarray}
&& \underbrace{\left[\begin{array}{c}
{\bf U}_3 \\
{\bf U}_4
\end{array}
\right]}_{2(m-d)\times 4(m-d)} {\bf V}_W = {\bf 0}, 
~\mbox{where for $e \in \{3,4\}$,}~\underbrace{[{\bf U}_e~-{\bf P}_e]}_{(m-d) \times (4m-d)} \label{eq:3p}
\left[\begin{array}{c}
{\bf V}\\
{\bf H}_e
\end{array}
\right]_{(4m-d)\times 3m} = {\bf 0}.
\end{eqnarray}
Security (\ref{eq:sec}) is guaranteed by (\ref{eq:3p}). For correctness (\ref{eq:corr}), we require that
\begin{eqnarray}
{\bf V}_W(1:2(m-d), :), {\bf V}_W(2(m-d)+1:4(m-d), :)~\mbox{both have full rank}
\end{eqnarray}
which is similarly proved by the Schwartz-Zippel lemma \cite{Demillo_Lipton, Schwartz, Zippel} and the property that the determinant polynomials (of variables from ${\bf H}_k, k \in [1:4]$) are not the zero polynomial. The following realization of ${\bf H}_k$ shows that the determinant polynomials are not always zero.
\begin{eqnarray}
&& {\bf H}_1(1:2(m-d),:) {\bf s} = s_{1:2(m-d)}, ~{\bf H}_2(1:2(m-d)),:) {\bf s} = s_{(2m-2d+1):4(m-d)} \\
&& {\bf H}_3 {\bf s} = (s_{1:(m-d)}+s_{(2m-2d+1):3(m-d)}; s_{(4m-4d+1):3m}) \\
&& {\bf H}_4 {\bf s}= (s_{(m-d+1):2(m-d)}+s_{(3m-3d+1):4(m-d)}; s_{(4m-4d+1):3m}) \\
&& \left[\begin{array}{c}
{\bf U}_3\\
{\bf U}_4
\end{array}\right] = [
{\bf I}_{2(m-d)\times 2(m-d)}~{\bf I}_{2(m-d)\times 2(m-d)}], 
~ {\bf V}_W = \left[ \begin{array}{c}
{\bf I}_{2(m-d) \times 2(m-d)} \\
-{\bf I}_{2(m-d)\times 2(m-d)}
\end{array}
\right].
\end{eqnarray}
The proof of achievability when $2 \leq \gamma \leq 5/2$ is complete.

\subsubsection{$3/2\leq \gamma \leq 5/3$}

Converse follows from (\ref{con:rate}) in Theorem \ref{thm:con}. $R \leq H(z_q|z_e) = H(z_q, z_e) - H(z_e) = m - d$ almost surely as $\gamma < 2$ so that from $z_q, z_e$, we can recover $s_{1:m}$ with high probability. 

For achievability, we note that there is a $(2d-m)$-dimensional overlap between ${\bf H}_{1}$ and ${\bf H}_2$ (the two qualified keys), 
\begin{eqnarray}
{\bf H}_{12} \triangleq {\bf V}_{12} {\bf H}_1, ~\mbox{where}~[{\bf V}_{12} ~-{\bf V}_{21}]_{(2d-m)\times 2d} \left[\begin{array}{c}
{\bf H}_{1} \\
{\bf H}_2
\end{array}
\right]_{2d \times m} = {\bf 0}_{(2d-m)\times m}
\end{eqnarray}
and similarly, for the two eavesdropping receivers, ${\bf H}_{3}$ and ${\bf H}_4$ have a $(2d-m)$-dimensional overlap, from which we will use $2m - 3d \leq 2d - m$ (recall that $\gamma \leq 5/3$) generic dimensions.
\begin{eqnarray}
{\bf H}_{34} \triangleq {\bf V}_{34} {\bf H}_3, ~\mbox{where}~[{\bf V}_{34} ~-{\bf V}_{43}]_{(2m-3d)\times 2d} \left[\begin{array}{c}
{\bf H}_{3} \\
{\bf H}_{4}
\end{array}
\right]_{2d \times m} = {\bf 0}_{(2m-3d)\times m}.
\end{eqnarray}
With high probability, ${\bf H}_1$ and ${\bf H}_2$ span the overall key space $s_{1:m}$ and we can express ${\bf H}_{34}$ as linear combinations of the rows of ${\bf H}_1$ and ${\bf H}_2$.
\begin{eqnarray}
&& {\bf H}_1\overset{\mbox{\scriptsize invertible}}{\longleftrightarrow} [{\bf H}_1(1:m-d, :); ~{\bf H}_{12}], 
~{\bf H}_2 \overset{\mbox{\scriptsize invertible}}{\longleftrightarrow} [{\bf H}_2(1:m-d, :); {\bf H}_{12}]; \\ 
&& {\bf H}_{34} = {\bf C}_{1} {\bf H}_1(1:m-d, :) + {\bf C}_2 {\bf H}_2(1:m-d, :) + {\bf C}_{12} {\bf H}_{12}
\end{eqnarray}
where ${\bf C}_{1}, {\bf C}_{2} \in \mathbb{F}_p^{(2m-3d) \times (m-d)}, {\bf C}_{12} \in \mathbb{F}_p^{(2m-3d) \times (2d-m)}$.

We wish to send $L_W = m-d$ message symbols $W$. Specifically, the first $2d-m$ message symbols are denoted by ${\bf W}_1\in \mathbb{F}_p^{(2d-m) \times 1}$ and last $2m-3d$ message symbols are denoted by ${\bf W}_2 \in \mathbb{F}_p^{(2m-3d) \times 1}$. The message $W$ is sent over $L=1$ key block and the transmit signal is set as
\begin{eqnarray}
X = \left(\begin{array}{c}
{\bf W}_1 + {\bf H}_{12} s_{1:m} \\
{\bf W}_2 + {\bf C}_{1} {\bf H}_1(1:m-d, :) s_{1:m} \\
- {\bf C}_{12} {\bf W}_1 - {\bf W}_2 + {\bf C}_{2} {\bf H}_2(1:m-d, :) s_{1:m}
\end{array}
\right) \in \mathbb{F}_p^{(3m-4d) \times 1}.
\end{eqnarray}
Correctness (\ref{eq:corr}) is easily seen, as qualified Receiver $1$ can obtain ${\bf W}_1, {\bf W}_2$ from the first two row blocks of $X$, and qualified Receiver $2$ can obtain ${\bf W}_1, {\bf W}_2$ from the first and third row block of $X$. Security (\ref{eq:sec}) holds because the overlap of each of $z_3, z_4$ (of dimension $d$) with the key space of $X$ (of dimension $3m-4d$) is ${\bf H}_{34}$ (of dimension $d + 3m-4d - m = 2m - 3d$), along which the projection of $W$ in $X$ is null. 
\begin{eqnarray}
\mbox{(Key space of $X$)} \cap \mbox{rowspace}({\bf H}_3) = \mbox{(Key space of $X$)} \cap \mbox{rowspace}({\bf H}_4) = {\bf H}_{34}.
\end{eqnarray}
Note that this design wherein both overlaps are the same space also follows from interference alignment principles. 
We need to ensure that the overlap is only ${\bf H}_{34}$ with high probability, i.e., the direct sum of the key space in $X$ and the row space of each of ${\bf H}_3, {\bf H}_4$ have full row rank almost surely, which is formalized by the Schwartz-Zippel lemma \cite{Demillo_Lipton, Schwartz, Zippel}. Note that the remaining rows of ${\bf H}_3$ and ${\bf H}_4$ (except ${\bf H}_{34}$) are generated independently of ${\bf H}_{1}, {\bf H}_2$, so the determinant polynomials of corresponding matrices contain distinct monomials and are thus non-zero.


\subsection{Proof of Theorem \ref{thm:newcon}: New Converse} \label{sec:newcon}
Following the insights from (\ref{eq:22in}), we consider the overlap of the transmit signal $X$ and each eavesdropping key spaces $Z_e, e \in [N+1:K]$ conditioned on $U_\mathcal{E}$. On the one hand, 
\begin{eqnarray}
I(X; Z_e | U_\mathcal{E}) &\overset{(\ref{sec})}{=}& I(X; Z_e, W | U_\mathcal{E}) \\ 
&=& I(X; Z_{1:K}, Z_e, W | U_\mathcal{E}) - I(X; Z_{1:K} | Z_e, W, U_\mathcal{E}) \\
&\geq& I(X; Z_{1:K}, W | U_\mathcal{E}) - H(Z_{1:K} | Z_e, W, U_\mathcal{E})\\
&\overset{(\ref{h2})}{=}& I(X; Z_{1:K}, W | U_\mathcal{E}) - H(Z_{1:K} | U_\mathcal{E}) + H(Z_e | U_\mathcal{E}) \\
&\overset{}{=}& I(X; Z_{1:K}, W | U_\mathcal{E}) - H(z_{1:K} | u_\mathcal{E}) L + H(z_e | u_\mathcal{E}) L . \label{eq:new1}
\end{eqnarray}
On the other hand,
\begin{eqnarray}
\sum_{e = N+1}^K I(X; Z_e | U_\mathcal{E}) &\leq& \sum_{e=N+1}^K I(X, Z_{N+1 : e-1}; Z_e | U_\mathcal{E})\\
&\overset{(\ref{eq:newind})}{=}& \sum_{e=N+1}^K I(X; Z_e | Z_{N+1 : e-1}, U_\mathcal{E}) \\
&=& I(X; Z_{N+1:K} | U_\mathcal{E}) \\
&\leq& I(X; Z_{1:K} | U_\mathcal{E}) \\
&=& I(X; Z_{1:K}, W | U_\mathcal{E}) - I(X; W | Z_{1:K}, U_\mathcal{E}) \\
&\overset{(\ref{corr})}{=}& I(X; Z_{1:K}, W| U_\mathcal{E}) - H(W | Z_{1:K}, U_\mathcal{E}) \\
&\overset{(\ref{h2})}{=}& I(X; Z_{1:K}, W | U_\mathcal{E}) - L_W. \label{eq:new2}
\end{eqnarray}
Adding (\ref{eq:new1}) for all $e \in [N+1:K]$ and combining with (\ref{eq:new2}), we have
\begin{eqnarray}
&& (K-N) \Big( I(X; Z_{1:K}, W | U_\mathcal{E}) - H(z_{1:K} | u_\mathcal{E}) L \Big) + \sum_{e = N+1}^K H(z_e | u_\mathcal{E}) L \notag\\
&&\leq I(X; Z_{1:K}, W | U_\mathcal{E}) - L_W \\
&\Rightarrow& L_W + (K-N-1) I(X; W, Z_{1:K} | U_\mathcal{E}) \leq (K-N) H(z_{1:K} | u_\mathcal{E}) L - \sum_{e = N+1}^K H(z_e | u_\mathcal{E}) L
\end{eqnarray}
and normalizing by $L$ gives us the desired bound.

\subsection{Proof of Theorem \ref{thm:nbig}: Asymptotic Alignment} \label{sec:nbig}
Adding more receivers (qualified or eavesdropping) cannot help so that for the converse proof, it suffices to consider the $N=2, E=2$ system. From Theorem \ref{thm:22}, we have the desired bound $\overline{R}(\gamma = 2) \leq 2/3$. Next, we provide the achievability proof, which is asymptotic, i.e., the normalized $\overline{R}(\gamma = 2)$ approaches $2/3$ when a parameter of the scheme goes to infinity.

\subsubsection{Achievability when $N > 2, E = 2$}
Suppose $2d/3 = \binom{\Delta+1}{2N}$ for some positive integer $\Delta$ (later $\Delta$ will be driven to infinity).
Define a matrix that is parameterized by $\Delta$ and is comprised of a collection of row vectors as follows.
\begin{eqnarray}
{\bf V}^\Delta = \left\{ {\bf 1} \left( \prod_{q_1, q_2} ({\bf H}_{q_1}^{[1]})^{\alpha_{q_1}^{[1]}} ({\bf H}_{q_2}^{[2]})^{\alpha_{q_2}^{[2]}} \right):
\sum_{q_1, q_2} (\alpha_{q_1}^{[1]} + \alpha_{q_2}^{[2]}) \leq \Delta, \alpha_{q_1}^{[1]}, \alpha_{q_2}^{[2]} \in \mathbb{Z}_+, q_1, q_2 \in [1:N]
\right\}
\end{eqnarray}
where ${\bf 1}$ is the $1\times d$ all $1$ row vector and ${\mathbb{Z}_+}$ is the set of positive integers. 
Thus ${\bf V}^{\Delta}$ contains product terms up to degree $\Delta$.
The number of row vectors in ${\bf V}^{\Delta+1}$ is equal to $\binom{\Delta+1}{2N}$, which has been set to $2d/3$. 

Set $L=1$, $L_W = \binom{\Delta}{2N} < 2d/3$, $L_X = 2\binom{\Delta+1}{2N} = 4d/3$ and the transmit signal $X$ as
\begin{eqnarray}
X = {\bf V}_W W + \left[\begin{array}{cc}
{\bf V}^{\Delta+1} & {\bf 0} \\
{\bf 0} & {\bf V}^{\Delta+1} 
\end{array} 
\right] s_{1:2d} = 
{\bf V}_W W + \left[\begin{array}{c}
{\bf V}^{\Delta+1} s_{1:d} \\
{\bf V}^{\Delta+1} s_{d+1:2d}
\end{array}
\right]
\end{eqnarray}
where $X \in \mathbb{F}_p^{L_X \times 1}, {\bf V}_W \in \mathbb{F}_p^{L_X \times L_W}, W \in \mathbb{F}_p^{L_W \times 1}$. ${\bf V}_W$ is designed so that
\begin{eqnarray}
\underbrace{\left[\begin{array}{c}
{\bf U}_{K-1} \\
{\bf U}_{K}
\end{array}
\right]}_{2d/3\times 4d/3} {\bf V}_W = {\bf 0}, 
~\mbox{where for $e \in \{K-1,K\}$,}~\underbrace{[{\bf U}_e~-{\bf P}_e]}_{d/3 \times 7d/3}
\underbrace{\left[\begin{array}{cc}
{\bf V}^{\Delta+1} & {\bf 0} \\
{\bf 0} & {\bf V}^{\Delta+1} \\
{\bf H}_e^{[1]} & {\bf H}_e^{[2]}
\end{array}
\right]}_{7d/3 \times 2d} = {\bf 0}. \label{eq:secp}
\end{eqnarray}
Note that $L_W < 2d/3$ so that ${\bf V}_W$ exists. Security (\ref{eq:sec}) follows from (\ref{eq:secp}) and ${\bf V}^{\Delta+1}$ is determined fully by qualified key matrices ${\bf H}_q^{[1]}, {\bf H}_q^{[2]}, q \in [1:N]$ thus ${\bf U}_e$ has $d/3$ rows almost surely. 
Correctness (\ref{eq:corr}) is due to the observation that $\forall q \in [1:N]$
\begin{eqnarray}
&& {\bf V}^\Delta {\bf H}_q^{[1]} \subset {\bf V}^{\Delta+1}, {\bf V}^\Delta {\bf H}_q^{[2]} \subset {\bf V}^{\Delta+1} \\
&\Rightarrow& {\bf V}^{\Delta} z_q = {\bf V}^{\Delta} {\bf H}_q^{[1]} s_{1:d} + {\bf V}^{\Delta} {\bf H}_q^{[2]} s_{d+1:2d} ~\mbox{($L_W$ rows)}\notag\\
&& \mbox{can be obtained from the rows of}~{\bf V}^{\Delta+1} s_{1:d}, {\bf V}^{\Delta+1} s_{d+1:2d} \notag\\
&& \mbox{i.e., there exists ${\bf D}_q \in \mathbb{F}_p^{L_W \times L_X}$ s.t.}~{\bf D}_q [{\bf V}^{\Delta+1} s_{1:d}; {\bf V}^{\Delta+1} s_{d+1:2d}] = {\bf V}^{\Delta} z_q \\
&\Rightarrow& {\bf D}_q X - {\bf V}^{\Delta} z_q = {\bf D}_q {\bf V}_W W
\end{eqnarray}
so we need to ensure ${\bf D}_q {\bf V}_W$ has full rank almost surely. The determinant polynomial of ${\bf D}_q {\bf V}_W$ is not always zero as there exists one such realization ${\bf H}_k, k \in [1:K]$.
Finally, the normalized rate achieved is
\begin{eqnarray}
\overline{R}(\gamma = 2) = \frac{L_W}{d} = \frac{\binom{\Delta}{2N}}{\frac{3}{2} \binom{\Delta+1}{2N}} = \frac{2}{3} \frac{\Delta+1 - 2N}{\Delta+1} \rightarrow \frac{2}{3} ~\mbox{as}~\Delta \rightarrow \infty.
\end{eqnarray}

\subsubsection{Achievability when $N = 2, E > 2$}\label{sec:eleak}
Suppose $d/3 = \binom{\Delta+1}{2E}$ for some positive integer $\Delta$.
Define a matrix that is comprised of the following row vectors. Note that $K = E+2$ and receivers $3$ to $K$ are eavesdroppers.
\begin{eqnarray}
{\bf V}^\Delta = \left\{ {\bf 1} \left( \prod_{e_1, e_2} ({\bf H}_{e_1}^{[1]})^{\alpha_{e_1}^{[1]}} ({\bf H}_{e_2}^{[2]})^{\alpha_{e_2}^{[2]}} \right):
\sum_{e_1, e_2} (\alpha_{e_1}^{[1]} + \alpha_{e_2}^{[2]}) \leq \Delta, \alpha_{e_1}^{[1]}, \alpha_{e_2}^{[2]} \in \mathbb{Z}_+, e_1, e_2 \in [3:K]
\right\}
\end{eqnarray}
where ${\bf 1}$ is the $1\times d$ all $1$ row vector and ${\bf V}^{\Delta}$ contains product terms up to degree $\Delta$. The number of row vectors in ${\bf V}^{\Delta+1}$ is equal to $\binom{\Delta+1}{2E}$, which has been set to $d/3$. 

We wish to design the scheme so that the overlap of each eavesdropping key and the key space of $X$ belongs to the space spanned by $[{\bf V}^{\Delta+1} s_{1:d}; ~{\bf V}^{\Delta+1} s_{d+1:2d}]$. To this end, we wish to see how to create this space from the qualified keys. As $\gamma = 2$, so the qualified keys $z_1, z_2$ are invertible to all $2d$ basis key symbols almost surely. Define ${\bf H}_{12} \triangleq [{\bf H}_1; ~{\bf H}_2], \overline{s}_{1:2d} \triangleq {\bf H}_{12} s_{1:2d}$, then $z_1 = \overline{s}_{1:d}, z_2 = \overline{s}_{d+1:2d}$. The $2d \times 2d$ square matrix ${\bf H}_{12}$ is invertible almost surely, i.e., $\mbox{inv}({\bf H}_{12})$ exists. We define the four $d\times d$ sub-matrix of $\mbox{inv}({\bf H}_{12})$ as follows,
\begin{eqnarray}
\mbox{inv}({\bf H}_{12}) = \left[
\begin{array}{cc}
\mbox{inv}^{[11]}({\bf H}_{12}) & \mbox{inv}^{[12]}({\bf H}_{12}) \\
\mbox{inv}^{[21]}({\bf H}_{12}) & \mbox{inv}^{[22]}({\bf H}_{12}) \\
\end{array}
\right] \in \mathbb{F}_p^{2d \times 2d}, \mbox{inv}^{[ij]}({\bf H}_{12}) \in \mathbb{F}_p^{d\times d}, i, j \in \{1,2\}
\end{eqnarray}
and then we have
\begin{eqnarray}
\left[\begin{array}{cc}
{\bf V}^{\Delta+1} s_{1:d} \\
{\bf V}^{\Delta+1} s_{d+1:2d}
\end{array} 
\right] &=&
\left[\begin{array}{cc}
{\bf V}^{\Delta+1} & {\bf 0} \\
{\bf 0} & {\bf V}^{\Delta+1} 
\end{array} 
\right] s_{1:2d} \\
&=& 
\left[\begin{array}{cc}
{\bf V}^{\Delta+1} & {\bf 0} \\
{\bf 0} & {\bf V}^{\Delta+1} 
\end{array} 
\right] 
\mbox{inv}({\bf H}_{12}) \overline{s}_{1:2d} \\
&=& \left[\begin{array}{cc}
{\bf V}^{\Delta+1} & {\bf 0} \\
{\bf 0} & {\bf V}^{\Delta+1} 
\end{array} 
\right] 
\left[
\begin{array}{cc}
\mbox{inv}^{[11]}({\bf H}_{12}) & \mbox{inv}^{[12]}({\bf H}_{12}) \\
\mbox{inv}^{[21]}({\bf H}_{12}) & \mbox{inv}^{[22]}({\bf H}_{12}) \\
\end{array}
\right]  
\overline{s}_{1:2d} \\
&=& \left[\begin{array}{cc}
{\bf V}^{\Delta+1} \mbox{inv}^{[11]}({\bf H}_{12})  z_1 + {\bf V}^{\Delta+1} \mbox{inv}^{[12]}({\bf H}_{12}) z_2 \\
{\bf V}^{\Delta+1} \mbox{inv}^{[21]}({\bf H}_{12}) z_1 + {\bf V}^{\Delta+1} \mbox{inv}^{[22]}({\bf H}_{12}) z_2
\end{array} 
\right] \label{eq:ee0}
\end{eqnarray}

Set $L=1$, $L_W = 2d/3$, $L_X = 4d/3$ and the transmit signal $X$ as
\begin{eqnarray}
X = {\bf V}_W W + {\bf V} s_{1:2d} = 
\left[\begin{array}{c}
W_{1:d/3} \\
W_{d/3+1:2d/3} \\
- W_{1:d/3} \\
- W_{d/3+1:2d/3} 
\end{array}
\right]
 + \left[\begin{array}{c}
{\bf V}^{\Delta+1} \mbox{inv}^{[11]}({\bf H}_{12})  z_1 \\
{\bf V}^{\Delta+1} \mbox{inv}^{[21]}({\bf H}_{12}) z_1 \\
 {\bf V}^{\Delta+1} \mbox{inv}^{[12]}({\bf H}_{12}) z_2 \\
 {\bf V}^{\Delta+1} \mbox{inv}^{[22]}({\bf H}_{12}) z_2
\end{array}
\right] \label{eq:ee1}
\end{eqnarray}
where $X \in \mathbb{F}_p^{L_X \times 1}, {\bf V}_W \in \mathbb{F}_p^{L_X \times L_W}, W \in \mathbb{F}_p^{L_W \times 1}, {\bf V} \in \mathbb{F}_p^{L_X \times 2d}$. 

We prove that the scheme is correct and the leakage is small compared to $d$. Correctness (\ref{eq:corr}) follows from the observation that from the first two row blocks of $X$, qualified Receiver $1$ can decode all $L_W = 2d/3$ symbols of $W$, and from the last two row blocks of $X$, qualified Receiver $2$ can decode all $L_W = 2d/3$ symbols of $W$. Next, consider security (\ref{eq:sec}). Note that $\forall e \in [3:K]$
\begin{eqnarray}
&& {\bf V}^\Delta {\bf H}_e^{[1]} \subset {\bf V}^{\Delta+1}, {\bf V}^\Delta {\bf H}_e^{[2]} \subset {\bf V}^{\Delta+1}\\
&\Rightarrow& \mbox{rowspan}([{\bf V}^\Delta {\bf H}_e^{[1]} ~~{\bf V}^\Delta {\bf H}_e^{[2]}]) \subset \mbox{rowspan}\left(
\left[\begin{array}{cc}
{\bf V}^{\Delta+1} & {\bf 0} \\
{\bf 0} & {\bf V}^{\Delta+1} 
\end{array} 
\right]
\right)\\
&\Rightarrow& {\bf V}^\Delta z_e =  {\bf V}^\Delta {\bf H}_e^{[1]} s_{1:d} + {\bf V}^\Delta {\bf H}_e^{[2]} s_{d+1:2d}
~\mbox{can be obtained from}~\notag \\
&&
\left[\begin{array}{cc}
{\bf V}^{\Delta+1} & {\bf 0} \\
{\bf 0} & {\bf V}^{\Delta+1} 
\end{array} 
\right] s_{1:2d} = 
\left[\begin{array}{cc}
{\bf V}^{\Delta+1} s_{1:d} \\
{\bf V}^{\Delta+1} s_{d+1:2d}
\end{array} 
\right] \label{eq:ee2}.
\end{eqnarray}
The row space of $z_e$ (of dimension of $d$) and the key space of $X$ (of dimension $4d/3$) have overlap of $d + 4d/3 - 2d = d/3$ dimensions with high probability (easily verified by showing that the determinant polynomials are non-zero). The row space of ${\bf V}^\Delta$ belongs to this overlap as it can be obtained from ${\bf V} s_{1:2d}$ (which follows from the design of the scheme, refer to (\ref{eq:ee0}), (\ref{eq:ee1}), (\ref{eq:ee2})). Further, the dimension of the row space of ${\bf V}^\Delta$ is equal to the number of row vectors in ${\bf V}^\Delta$ with high probability, which is $\binom{\Delta}{2E}$. The projection of the message $W$ in the transmit signal $X$ to this overlapping space ${\bf V}^\Delta z_e$ is zero thus nothing is revealed. Except from (orthogonal to) the row space of ${\bf V}^\Delta$, the remaining overlap of the row space of $z_e$ and the key space of $X$ has dimension at most
\begin{eqnarray}
&& d/3 - \dim\left(\mbox{rowspace}({\bf V}^\Delta) \right) = \binom{\Delta+1}{2E} - \binom{\Delta}{2E} \\
&\Rightarrow& \frac{d/3 - \dim\left(\mbox{rowspace}({\bf V}^\Delta) \right)}{d} = \frac{1}{3} \left( 1 - \frac{\binom{\Delta}{2E}}{\binom{\Delta+1}{2E}} \right) = \frac{2E}{3(\Delta+1)} \rightarrow 0 ~\mbox{as $\Delta,d$}~\rightarrow \infty.
\end{eqnarray}
Therefore, the leakage vanishes with the spatial dimension $d$. As we allow $d$ to approach infinity, the normalized leakage is negligible, i.e., $I(W; X, Z_e) = o(d)$ (the derivation is the $\epsilon$ leakage relaxation of the zero leakage counterpart in (\ref{eq:e0})).

\section{Conclusion}
In this work, we show that for the secure groupcast problem which involves no noise, the communicate rate is not fully specified by the source variables in the problem statement. As a result, a more general entropic description that includes auxiliary variables for the achievability and converse is required. Additional insights are necessary to reveal the structure of auxiliary variables. 

We also study the generic secure groupcast problem where each key is comprised of a number of generic linear combinations. The groupcast rate is measured as a function of the ratio of the dimension of the overall key space to the dimension of each receiver's key space. The feasibility of linear schemes is stated in terms of space projections and overlaps, which leads to the natural application of various interference alignment schemes originated in wireless communications. While complete answers are obtained when the overall key space is either large or small, the intermediate cases are open and call for more advanced techniques.

\let\url\nolinkurl
\bibliographystyle{IEEEtran}
\bibliography{Thesis}
\end{document}